\documentclass[aps,pre,twocolumn,superscriptaddress,showkeys,showpacs,amsfonts,amssymb,amsmath]{revtex4}

\usepackage{bm}
\usepackage{graphicx}
\usepackage{subfigure} 

\begin{document}

\newcommand{\optColourPrefix}[0]{}
\newcommand{\maxMagColour}[0]{black}
\newcommand{\optColourCaptionPrefix}[0]{}


\title{Local information transfer as a spatiotemporal filter for complex systems}


\author{Joseph T. Lizier}
 \email[]{jlizier@it.usyd.edu.au}
\affiliation{CSIRO Information and Communications Technology Centre,
Locked Bag 17, North Ryde, NSW 1670, Australia}
\affiliation{School of Information Technologies, The University of Sydney,
NSW 2006, Australia}

\author{Mikhail Prokopenko}
\affiliation{CSIRO Information and Communications Technology Centre,
Locked Bag 17, North Ryde, NSW 1670, Australia}

\author{Albert Y. Zomaya}
\affiliation{School of Information Technologies, The University of Sydney,
NSW 2006, Australia}


\date{\today}

\begin{abstract}
We present a measure of local information transfer, derived from an existing averaged information-theoretical measure, namely transfer entropy. Local transfer entropy is used to produce profiles of the information transfer into each spatiotemporal point in a complex system. These spatiotemporal profiles are useful not only as an analytical tool, but also allow explicit investigation of different parameter settings and forms of the transfer entropy metric itself. As an example, local transfer entropy is applied to cellular automata, where it is demonstrated to be a novel method of filtering for coherent structure. More importantly, local transfer entropy provides the first quantitative evidence for the long-held conjecture that the emergent traveling coherent structures known as particles (both gliders and domain walls, which have analogues in many physical processes) are the dominant information transfer agents in cellular automata.
\end{abstract}


\maketitle


\section{\label{intro}Introduction}
Information transfer is widely considered to be a vital component of complex nonlinear behavior in spatiotemporal systems, for example in: particles in cellular automata (CAs) \cite{mitch94,mitch96,mitch98,hord01,lang90,wue99,wolf84b}, self-organization caused by dipole-dipole interactions in microtubules \cite{brown99}, soliton dynamics and collisions \cite{edm93}, wave-fragment propagation in Belousov-Zhabotinsky media \cite{send01}, solid-state phase transitions in crystals \cite{varn04}, influence of intelligent agents over their environments \cite{kly05b}, and inducing emergent neural structure \cite{lung06}.
The very nature of information transfer in complex systems is a popular topic itself, for example in the conflicting suggestions that information transfer is maximized in complex dynamics \cite{sole01,mira95}, or alternatively at an intermediate level with maximization leading to chaos \cite{lang90,coff98}.
Yet while the literature contains many measures of complexity (e.g. \cite{sha04,wue99}), quantitative studies of information transfer are comparatively absent.

Information transfer is popularly understood in terms of the aforementioned recognized instances, which suggest a \textit{directional} signal or communication of \textit{dynamic} information between a source and receiver. Defining information transfer as the dependence of the next state of the receiver on the previous state of the source \cite{jaku97} is typical, though it is incomplete according to Schreiber's criteria \cite{schr00} requiring the definition to be both directional and dynamic. In this paper, we accept Schreiber's definition \cite{schr00} of (predictive) information transfer as the average information contained in the source about the next state of the destination that was not already contained in the destination's past. This definition results in the measure for information transfer known as \textit{transfer entropy} \cite{schr00}, quantifying ``the statistical coherence between systems evolving in time" in a directional and dynamic manner.

We derive a measure of \textit{local} information transfer from this existing averaged information-theoretical measure, transfer entropy. Local transfer entropy characterizes the information transfer into each spatiotemporal point in a given system as opposed to a global average over all points in an information channel. Local metrics within a global average are known to provide important insights into the dynamics of nonlinear systems \cite{das02}: here, the local transfer entropy provides spatiotemporal profiles of information transfer, useful analytically in highlighting or filtering ``hot-spots" in the information channels of the system. The local transfer entropy also facilitates close study of different forms and parameters of the averaged metric, in particular the importance of conditioning on the past history of the information destination, and the possibility of conditioning on other information sources. Importantly, through these applications the local transfer entropy provides insights that the averaged transfer entropy cannot.

We apply local transfer entropy to cellular automata (CAs): discrete dynamical systems consisting of an array of cells which each synchronously update their state as a function of the states of a fixed number of spatially neighboring cells using a uniform rule.
CAs are a classic example of complex behavior, and have been used to model a wide variety of real world phenomena
(see \cite{mitch98}). In particular, we examine \textit{elementary CAs} (\textit{ECAs}): 1D CAs using binary states, deterministic rules and one neighbor on either side (i.e. cell range $r = 1$). (For more complete definitions, including that of the Wolfram rule number convention for describing update rules, see \cite{wolf02}).

CAs are selected for experimentation here because they have been the subject of a large body of work regarding the qualitative nature of information transfer in complex systems \cite{lang90,mitch94,mitch96,mitch98,hord01,wue99,wolf84b}. As will be described here, there are well-known spatiotemporal structures in CAs which are qualitatively widely-accepted as being information transfer agents; this provides us with a useful basis for interpreting the quantitative results of our application.
The aforementioned studies revolve around emergent structure in CAs: \textit{particles}, \textit{gliders} and \textit{domains}.
A domain may be understood as a set of background configurations in a CA, any of which will update to another such configuration in the absence of a disturbance. Domains are formally defined within the framework of computational mechanics \cite{han92} as spatial process languages in the CA.
Particles are qualitatively considered to be moving elements of coherent spatiotemporal structure, in contrast to a background domain (see \cite{sha06} for a discussion of the term ``coherent structure" referring to particles in this context). Gliders are particles which repeat periodically in time while moving spatially (repetitive non-moving structures are known as \textit{blinkers}).
Formally, particles are defined as a boundary between two domains \cite{han92}; as such, they can also be termed as \textit{domain walls}, though this is typically used with reference to aperiodic particles.
It is widely suggested that particles form the basis of information transmission, since they appear to facilitate communication about the dynamics in one area of the CA to another area (e.g. \cite{lang90}). Furthermore, their interactions or collisions are suggested to form the basis of information modification, since the collisions appear to combine the communications in some decision process about the dynamics. In particular, these metaphor are found in studies of Turing universal computation with particles used to facilitate the transfer of information between processing elements (e.g. Conway's Game of Life \cite{con82} and see general discussion in \cite{mitch98}); analyses of CAs performing intrinsic, universal or other specific computation \cite{han92,han97,mitch94,mitch96}; studies of the nature of particles and their interactions (i.e. collisions) \cite{mitch94,hord01}; and attempts to automatically identify CA rules which give rise to particles, e.g. \cite{wue99,epp02}, suggesting these to be the most interesting and complex CA rules.
Despite such interest, no study has quantified the information transfer on average within specific channels or at specific spatiotemporal points in a CA, nor quantitatively demonstrated that particles (either in general, or gliders or domain walls as sub-classes) are in fact information transfer agents. (A rudimentary attempt was made via mutual information in \cite{lang90}, however we show that this is a symmetric measure not capturing directional transfer).

We hypothesize that \textit{application of a measure of local information transfer into each spatiotemporal point in CAs would reveal particles as the dominant information transfer agents}. Our results would have wide-ranging implications for the real-world systems mentioned earlier, given the power of CAs as model systems of the real world and the obvious analogy between particles in CAs and coherent spatiotemporal structures and hypothesized information transfer agents in other systems (e.g. known analogues of particles in physical processes such as pattern formation and solitons \cite{hord01,park86}; also waves of conformational change are said to perform signaling in microtubules \cite{brown99}). Where no CA model exists for a given system, our presentation of local transfer entropy is generic enough to still be directly applicable for investigation of that system, guided by the method of application to CAs.

Finally, several methods already exist for filtering the important structural elements in CAs \cite{han92,wue99,hel04,sha06}, which provide another important basis for comparison of our spatiotemporal local information transfer profiles (which can also be viewed as a method of filtering).
These methods include: finite state transducers to recognize the regular spatial language of the CA \cite{han92,han97}; local information (i.e. local spatial entropy rate) \cite{hel04}; displaying executing rules with the most frequently occurring rules filtered out \cite{wue99}; and local statistical complexity and local sensitivity \cite{sha06}. All of these successfully highlight particles. Hence, filtering is not a new concept; however the ability to filter for information transfer could provide the first thoroughly quantitative evidence that particles are the information transfer elements in CAs. Additionally, it would provide insight into information transfer in each specific channel or direction in the CA allowing more refined investigation than the single measures of other methods, and should reveal interesting differences in the parts of the structures highlighted.

We begin by providing background on required information-theoretical concepts, and subsequently introduce transfer entropy and derive the local transfer entropy from it. We also derive two distinct forms of the transfer entropy, namely apparent and complete, to be studied from a local viewpoint. The local transfer entropy is then applied to ECAs, highlighting particles (both gliders and domain walls) as expected, and so providing the first quantitative evidence for the widely-accepted conjecture that these are the dominant information transfer entities in CAs. The profiles also provide insights into the parameters and forms of the transfer entropy that its average is shown to be incapable of producing. We conclude with a summary of the important findings, compare our spatiotemporal profiles to other CA filtering methods, and describe further investigations we intend to perform with this metric.

\section{\label{info}Information-theoretical quantities}
Information theory (e.g. see \cite{mac03}) has proved to be a useful framework for the design and analysis of complex self-organized systems (for example, see an overview in \cite{pro06c} and specific examples in \cite{sha04,wue99,pro06b,kly05b,lung06}). This success, in addition to the highly abstract nature of information theory (which renders it portable between different types of complex systems), and its general ease of use, are reasons underlying its position as a leading framework for the analysis and design of complex systems.

The fundamental quantity is the \textit{Shannon entropy}, which represents the uncertainty associated with any measurement $x$ of a random variable \textit{X} (logarithms are in base 2, giving units in bits): $H(X) = -\sum_{x} p(x) \log{p(x)}$.
%
The \textit{conditional entropy} of \textit{X} given \textit{Y} is the average uncertainty that remains about $x$ when $y$ is known: $H(X|Y) = -\sum_{x,y} p(x,y) \log{p(x|y)}$.
The \textit{mutual information} between \textit{X} and \textit{Y} measures the average reduction in uncertainty about $x$ that results from learning the value of $y$, or vice versa:
\begin{subequations}
\begin{eqnarray}
	I(X;Y) = \sum_{x,y} p(x,y) \log{\frac{p(x,y)}{p(x)p(y)}}
	\label{eq:mi}, \\
	I(X;Y) = H(X)-H(X|Y) = H(Y)-H(Y|X)
	\label{eq:mi2}.
\end{eqnarray}
\end{subequations}

The \textit{conditional mutual information} between \textit{X} and \textit{Y} given \textit{Z} is the mutual information between \textit{X} and \textit{Y} when \textit{Z} is known: 
\begin{equation}
	I(X;Y|Z) = H(X|Z)-H(X|Y,Z)
	\label{eq:condMI}.
\end{equation}

The entropy rate (denoted as $h_\mu$) \cite{crutch03} is the limiting value of the conditional entropy of the next state $x_{n+1}$ of \textit{X} given knowledge of the previous $k-1$ states $x^{(k-1)}_{n}$ (up to and including time \textit{n}, i.e. $x_{n-k+2}$ to $x_n$) of \textit{X}:
\begin{eqnarray}
	h_\mu = \lim_{k \rightarrow \infty}{H ( x_{n+1} | x^{(k-1)}_{n})} 
	\label{eq:entratecond}.
\end{eqnarray}

\section{\label{localTE}Local Information Transfer}
It is natural to look to information theory for the concept of information transfer. As such, we adopt transfer entropy from this realm and subsequently derive local transfer entropy from it. Additionally, we provide comment on the parameters of the transfer entropy, and present the concepts of apparent and complete transfer entropy, and self-information transfer.

\subsection{\label{transferEntropy}Transfer Entropy}
As alluded to earlier, mutual information has been something of a de facto measure for information transfer in complex systems (e.g. \cite{wue99,sole01,sole04}). A major problem however is that mutual information contains no inherent directionality. Attempts to address this include using the previous state of the ``source" variable and the next state of the ``destination" variable (known as \textit{time-lagged mutual information}). However, Schreiber \cite{schr00} points out that this ignores the more fundamental problem that mutual information measures the \textit{statically} shared information between the two elements. (The same criticism applies to equivalent non information-theoretical definitions such as that in \cite{jaku97}).

To address these inadequacies Schreiber introduced \textit{transfer entropy} \cite{schr00}, the deviation from independence (in bits) of the state transition (from the previous state to the next state) of an information destination \textit{X} from the (previous) state of an information source \textit{Y}:
\begin{equation}
	T_{Y \rightarrow X} = \sum_{u_n}
    p(u_n)
    \log{ \frac{ p(x_{n+1}|x^{(k)}_{n},y^{(l)}_{n})}{p(x_{n+1}|x^{(k)}_{n})}}
	\label{eq:te}.
\end{equation}
Here \textit{n} is a time index, $u_n$ represents the state transition tuple $(x_{n+1},x^{(k)}_{n},y^{(l)}_{n})$, $x^{(k)}_{n}$ and $y^{(l)}_{n}$ represent the \textit{k} and \textit{l} past values of \textit{x} and \textit{y} up to and including time \textit{n} (with $k,l = 1$ being default choices).
Schreiber points out that this formulation of the transfer entropy is a truly dynamic measure, as a generalization of the entropy rate to more than one element to form a mutual information \textit{rate}.
The transfer entropy can be viewed as a \textit{conditional} mutual information \cite{ay06} (see Eq.~(\ref{eq:condMI})), casting it as the average information contained in the source about the next state $X'$ of the destination that was not already contained in the destination's past:
\begin{equation}
	T_{Y \rightarrow X} = I(Y;X'|X) = H(X'|X) - H(X'|X,Y)
	\label{eq:teCond}.
\end{equation}
This could be interpreted (following \cite{pro06c} and \cite{sole04}) as the diversity of state transitions in the destination minus assortative noise between those state transitions and the state of the source.
Importantly, as an information theoretic measure based on observational probabilities, the transfer entropy is applicable to both deterministic and stochastic systems.

Transfer entropy has been used to characterize information flow in sensorimotor networks \cite{lung06} and with respect to information closure \cite{bert06a} in two recent studies. We note the alternative perturbation-based candidate \textit{information flow} for quantifying information transfer from the perspective of causality rather than prediction; we intend to compare transfer entropy to this measure in future work. Furthermore, a separate notion of information flow in CAs was introduced in \cite{hel04} (connected to the local information though not used for filtering). There are several fundamental problems with this formulation however: it is only applicable to reversible CAs, only has meaning as information flow for deterministic mechanics, and is not able to distinguish information flow any more finely than information from the left and the right.

In this paper, we accept Schreiber's formulation of transfer entropy (Eq.~(\ref{eq:te})) as a theoretically correct quantitative definition of information transfer, from a predictive or computational perspective. However, this quantitative definition has not yet been unified with the accepted specific instances of information transfer (e.g. particles in CAs); these instances are local in space and time and to be investigated require a local measure of information transfer. In presenting local transfer entropy here, we seek to unify the apparently correct quantitative formulation of information transfer (i.e. transfer entropy) with accepted specific instances of information transfer.

\subsection{\label{localTESection}Local Transfer Entropy}
To derive a local transfer entropy measure, we first note that Eq.~(\ref{eq:te}) is summed over all possible state transition tuples $u_n = (x_{n+1},x^{(k)}_{n},y^{(l)}_{n})$, weighted by the probability of observing each such tuple. This probability $p(u_n)$ is operationally equivalent to the ratio of the count of observations $c(u_n)$ of $u_n$, to the total number of observations \textit{N} made: $p(u_n) = c(u_n) / N$. To precisely compute this probability, the ratio should be composed over all realizations of the observed variables (as described in \cite{sha01a}); realistically however, estimates will be made from a finite number of observations.
Subsequently, we replace the count by its definition $c(u_n) = \sum^{c(u_n)}_{a=1} 1$, leaving the substitution
$p(u_n) = \left(\sum^{c(u_n)}_{a=1} 1\right) / N$
into Eq.~(\ref{eq:te}):
\begin{equation}
	T_{Y \rightarrow X} = \frac{1}{N} \sum_{u_n}
    \left(\sum^{c(u_n)}_{a=1} 1\right)
    \log{ \frac{ p(x_{n+1}|x^{(k)}_{n},y^{(l)}_{n})}{p(x_{n+1}|x^{(k)}_{n})}}
	\label{eq:teDoubleSum}.
\end{equation}
The log term may then be brought inside this inner sum:
\begin{equation}
	T_{Y \rightarrow X} = \frac{1}{N} \sum_{u_n}
    \sum^{c(u_n)}_{a=1}
    \log{ \frac{ p(x_{n+1}|x^{(k)}_{n},y^{(l)}_{n})}{p(x_{n+1}|x^{(k)}_{n})}}
	\label{eq:teDoubleSum2}.
\end{equation}

This leaves a double sum running over each actual observation \textit{a} for each possible tuple observation $u_n$, which is equivalent to a single sum over all \textit{N} observations:
\begin{equation}
	T_{Y \rightarrow X} = \frac{1}{N} \sum^{N}_{n=1}
    \log{ \frac{ p(x_{n+1}|x^{(k)}_{n},y^{(l)}_{n})}{p(x_{n+1}|x^{(k)}_{n})}}
    \label{eq:te_observations}.
\end{equation}

It is clear then that the transfer entropy metric is an global \textit{average} (or expectation value) of a \textit{local transfer entropy} at each observation:
\begin{subequations}
\label{te_localFirst}
\begin{eqnarray}
	T_{Y \rightarrow X} = \left\langle t_{Y \rightarrow X}(n+1,k,l) \right\rangle
    \label{eq:te_expectation}; \\
  t_{Y \rightarrow X}(n+1,k,l) = 
  	\log{ \frac{ p(x_{n+1}|x^{(k)}_{n},y^{(l)}_{n})}{p(x_{n+1}|x^{(k)}_{n})}}
	  \label{eq:te_local1}.
\end{eqnarray}
\end{subequations}

The measure is \textit{local} in that it is defined at each time $n$ for each destination element $X$ in the system and each causal information source $Y$ of the destination. This method of forming a local information-theoretic measure by extracting the log term from a globally averaged measure is used less explicitly for the local excess entropy\cite{sha01a}, the local statistical complexity \cite{sha01a,sha06}, and the local information \cite{hel04}. It is applicable to any such information-theoretic metric: we form the \textit{local (time-lagged) mutual information} between the source and destination variables from Eq.~(\ref{eq:mi}) as:
\begin{equation}
	m(y^{(l)}_{n} ; x_{n+1}) = \log{\frac{p(y^{(l)}_{n},x_{n+1})}{p(x_{n+1})p(y^{(l)}_{n})}}
    \label{eq:miLocal},
\end{equation}
and similarly rewrite Eq.~(\ref{eq:teCond}) as the expectation value of a \textit{local conditional mutual information}: $T_{Y \rightarrow X} = \left\langle m(y^{(l)}_{n} ; x_{n+1} | x^{(k)}_{n}) \right\rangle$.

\begin{figure}
	\resizebox{85mm}{!}{\includegraphics{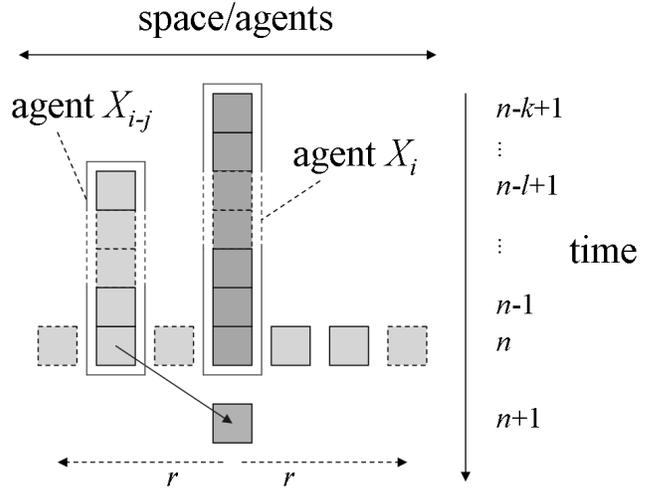}}
	\caption{\label{fig:localTeFigure}Local transfer entropy $t(i,j,n+1,k,l)$ is the information transfered from an $l$ sized block of the source cell $X_{i-j}$ to the destination cell $X_i$ at time step $n+1$, conditioned on $k$ past states of the destination cell. Note: $|j| \leq r$ for CAs.}
\end{figure}

For lattice systems such as CAs with \textit{spatially-ordered} sources and destinations, we represent the local transfer entropy to cell $X_{i}$ from $X_{i-j}$ at time $n+1$ as:
\begin{eqnarray}
	t(i,j,n+1,k,l) = \log{ \frac{ p(x_{i,n+1}|x^{(k)}_{i,n},x^{(l)}_{i-j,n})}{p(x_{i,n+1}|x^{(k)}_{i,n})}}
	\label{eq:teLocal}.
\end{eqnarray}
Similarly, the local (time-lagged) mutual information can be represented as: $m(i,j,n+1,l) = m(x^{(l)}_{i-j,n};x_{n+1})$.
Fig.~\ref{fig:localTeFigure} shows the local transfer entropy in a spatiotemporal system.
The metrics are defined for every spatiotemporal destination $(i,n)$, forming a \textit{spatiotemporal profile} for every information channel or direction \textit{j} where sensible values for CAs are within the cell range, $|j| \leq r$. Notice that $j$ represents the number of cells from the source to the destination, e.g. $j=1$ denotes transfer across one cell to the right per unit time step.
We use $T(j,k,l)$ to represent the average over all spatiotemporal points on the lattice.

Importantly, note that the destination's own historical values can indirectly influence it via the source, which may be mistaken as an independent flow of information from the source. This is only possible in systems such as CAs with \textit{bidirectional} information transfer. Such self-influence is a \textit{non-traveling} form of information (in the same way as standing waves are to energy); it is essentially static and can be viewed as the trivial part of information transfer. This non-traveling information is eliminated from the measurement by conditioning on the destination's history $x^{(k)}_{i,n}$.
Yet any self-influence transmitted prior to these \textit{k} values will not be eliminated; we generalize comments on the entropy rate in \cite{schr00} to suggest that taking the asymptote $k \rightarrow \infty$ is most correct for agents displaying non-Markovian dynamics (when considering their time-series in isolation). As such, we formalize the local transfer entropy as:
\begin{eqnarray}
	t(i,j,n+1,l) = \lim_{k \rightarrow \infty}{\log{ \frac{  p(x_{i,n+1}|x^{(k)}_{i,n},x^{(l)}_{i-j,n})}{p(x_{i,n+1}|x^{(k)}_{i,n})}}}
	\label{eq:teLocalLimitOrdered},
\end{eqnarray}
and similarly $t_{Y \rightarrow X}(n+1,l) = \lim_{k \rightarrow \infty}{t_{Y \rightarrow X}(n+1,k,l)}$
for a single source-destination pair. Computation at this limit is not feasible in general, so we retain $t_{Y \rightarrow X}(n,k,l)$ and $t(i,j,n,k,l)$ for estimation with finite $k$.

Also, we drop $l$ from the notation (e.g. $t(i,j,n)$ and $t(i,j,n,k)$) where the default setting of $l=1$ is used to measure transfer from the single previous state only.

\subsection{\label{completeAndApparent}Complete and Apparent Transfer Entropy}
The averaged transfer entropy is constrained between 0 and $\log{b}$ bits (where $b$ is the number of possible states for a discrete system): as a conditional mutual information, it can be either larger \textit{or} smaller than the corresponding mutual information \cite{mac03}. The \textit{local} transfer entropy however is not constrained so long as it averages into this range: it can be greater than $\log{b}$ for a significant local information transfer, and can also in fact be measured to be negative. Local transfer entropy is negative where (in the context of the history of the destination) the probability of observing the actual next state of the destination given the value of the source $p(x_{i,n+1}|x^{(k)}_{i,n},x^{(l)}_{i-j,n})$, is lower than that of observing that actual next state independently of the source $p(x_{i,n+1}|x^{(k)}_{i,n})$. In this case, the source element is actually \textit{misleading} about the state transition of the destination. It is possible for the source to be misleading in this context where other causal information sources influence the destination, or in a stochastic system. (Similarly a local mutual information, Eq.~(\ref{eq:miLocal}), can be negative).

Importantly, the transfer entropy may be conditioned on other possible information sources $Z$ \cite{schr00} (becoming $I(Y;X'|X,Z)$), to eliminate their influence from being mistaken as that of the source $Y$. To be explicit, we label calculations conditioned on no other information contributors (e.g. Eq.~(\ref{eq:teLocalLimitOrdered})) as \textit{apparent} transfer entropy.

For ECAs, conditioning on other possible information sources logically means conditioning on the other cells in the destination's neighborhood, which we know to be causal information contributors. Firstly, we represent the joint values of the neighborhood of the destination $x_{i,n+1}$, excluding the source for the transfer entropy calculation $x_{i-j,n}$ and the previous value of the destination $x_{i,n}$, as:
\begin{equation}
	v^{r}_{i,j,n} = \left\{ x_{i+q,n} | \forall q: -r \leq q \leq +r, q \neq -j, 0 \right\}
	\label{eq:neighborhood},
\end{equation}
where $r$ is the range of causal information contributors (i.e. the cell range for CAs).
We then derive local \textit{complete} transfer entropy as the information contained in the source about the next state of the destination that was not contained in the destination's past \textit{or} in other causal information sources $v^{r}_{i,j,n}$:
\begin{eqnarray}
	t_{c}(i,j,n+1)  = \lim_{k \rightarrow \infty}{\log{ \frac
		{p ( x_{i,n+1}|x^{(k)}_{i,n},x_{i-j,n},v^{r}_{i,j,n} )}
		{p ( x_{i,n+1}|x^{(k)}_{i,n},v^{r}_{i,j,n} )}}}
	\label{eq:teLocalComplete}.
\end{eqnarray}
Again, $t_{c}(i,j,n,k)$ denotes finite $k$ estimates. Eq.~(\ref{eq:teLocalComplete}) specifically considers systems where only immediately previous source values can be causal information contributors: here under complete conditioning $l > 1$ cannot add any information to the source. In other systems Eq.~(\ref{eq:teLocalComplete}) could be adjusted accordingly.
In deterministic systems (e.g. ECAs), complete conditioning renders $t_{c}(i,j,n) \geq 0$: it is not possible for the information source to be misleading when all other causal information sources are being considered. $T_c(j)$ represents the average over all spatiotemporal points on the lattice. Complete transfer entropy can be constructed for any system by conditioning out all causal information contributors apart from the information source under consideration.

\subsection{\label{summedProfile}Summed Information Transfer Profiles}
We label the case $j = 0$ as \textit{self-information transfer}, where the ``source" is the immediate past value of the destination. We condition this calculation on the \textit{k} values before the \textit{l} source values so as not to condition on the source. Self-information transfer computes the information contributed by the previous state of the given cell about its next state that was not contained in its prior history; this can be thought of as traveling information with an instantaneous velocity of zero. This is not a particularly useful quantity in and of itself, however
it helps to form a useful profile with transfer entropies for $j \neq 0$ in the \textit{summed local information transfer profiles}. These are defined for apparent and complete transfer entropy respectively as:
\begin{subequations}
\label{eq:teSummed}
\begin{eqnarray}
	t_{s}(i,n,k,l) = \sum^{r}_{j=-r} t(i,j,n,k,l)
	\label{eq:teSummedApparent}, \\
	t_{sc}(i,n,k) = \sum^{r}_{j=-r} t_c(i,j,n,k)
	\label{eq:teSummedComplete}.
\end{eqnarray}
\end{subequations}

\section{\label{results}Results and Discussion}

The local transfer entropy metrics were studied with several important ECA rules. We investigate the variation of the profiles as a function of $k$, examine the changing nature of the profiles with ECA type, and compare the apparent and complete metrics. Each instance was run from an initial randomized state of 10 000 cells, with the first 30 time steps eliminated to allow the CA to settle, and a further 600 time steps captured for investigation. All results were confirmed by at least 10 runs from different initial randomized states, and periodic boundary conditions were used. We fixed \textit{l} at 1: values of $l > 1$ are irrelevant for the complete metric when applied to CAs, and for the apparent metric we are interested in information directly transfered at the given local time step only.
For spatially-ordered systems with \textit{homogeneous} agents such as CAs, it is appropriate to estimate the probability distributions from all spatiotemporal observations (i.e. from the whole CA) of the corresponding channel rather than only the source-destination pair under measurement. 

We concentrate on rule 110 (a complex rule with several configurations of regular particles, or gliders) and rule 18 (a chaotic rule with irregular particles, or domain walls); the rule classification here is from \cite{wolf02}. The selection of these particular rules allow comparison with the results of other filtering techniques. We expect local information transfer profiles to highlight both regular and irregular particles, the important elements of structure in CAs which are conjectured to be the information transfer agents.

\subsection{\label{baseCases}Base comparison cases}
\begin{figure}
	\subfigure[]{\fbox{\label{fig:110raw-base}\includegraphics[width=0.23\textwidth]{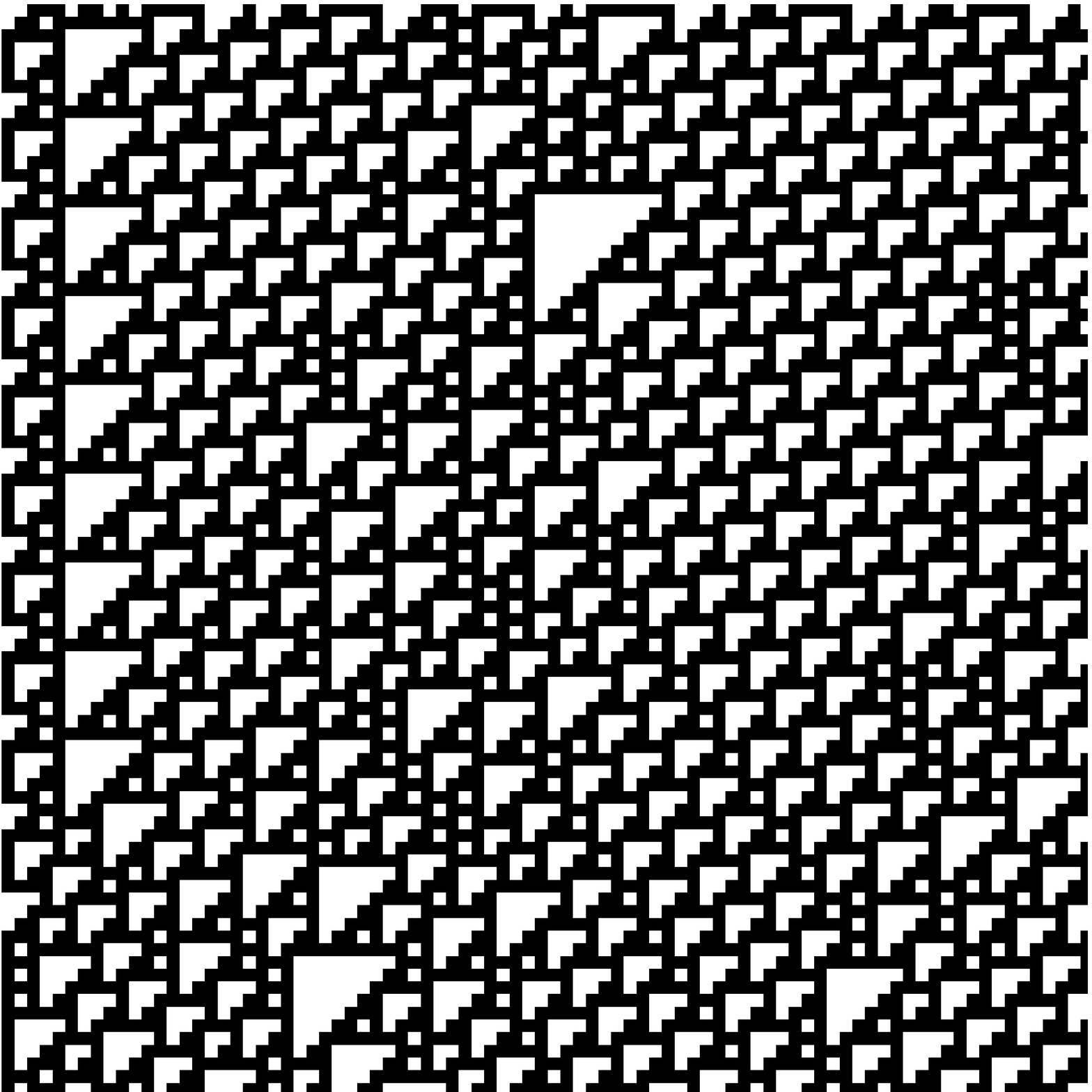}}}
	\subfigure[]{\fbox{\label{fig:110mi1-pos}\includegraphics[width=0.23\textwidth]{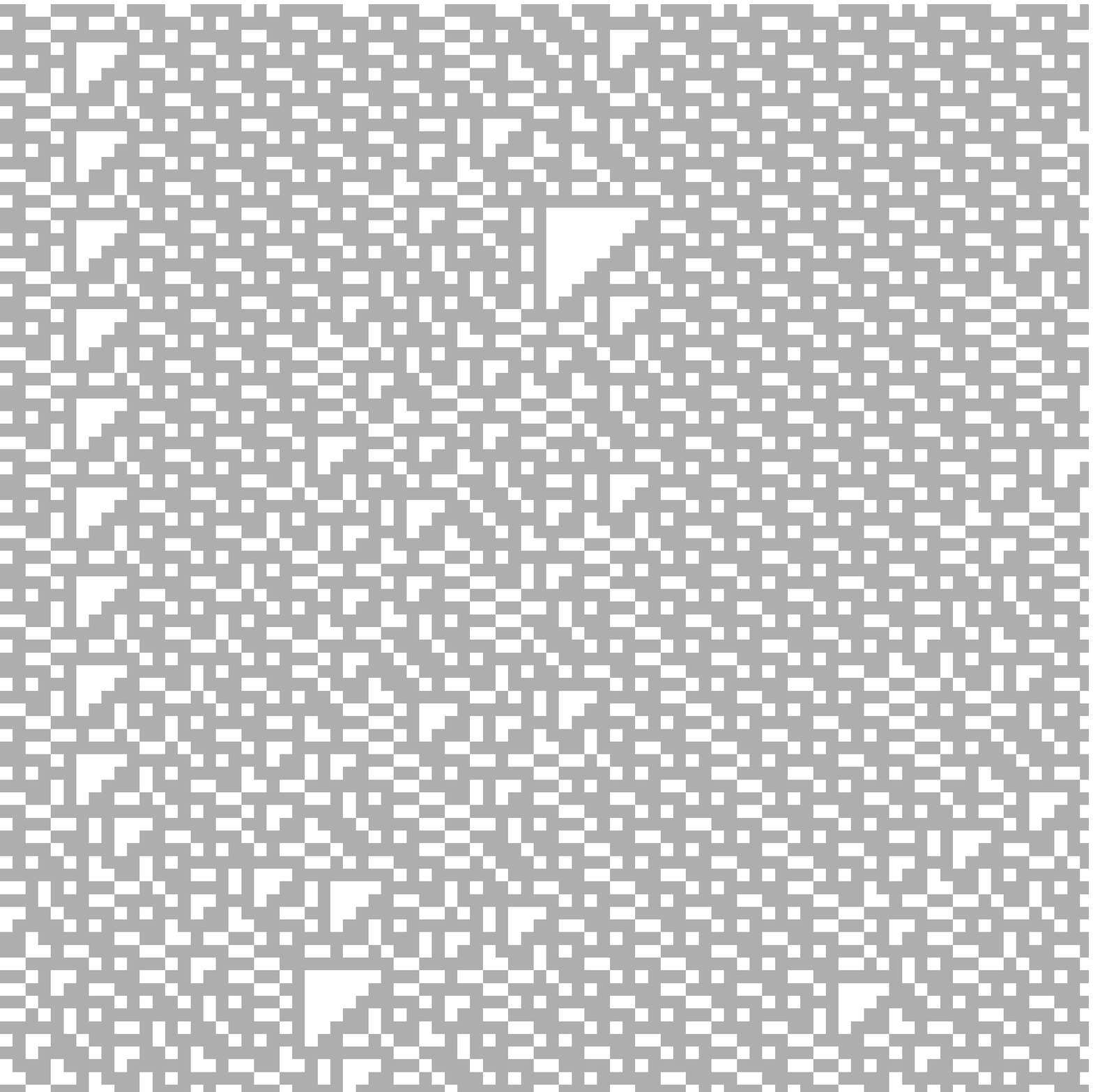}}}
	\subfigure[]{\fbox{\label{fig:110teComp1-1}\includegraphics[width=0.23\textwidth]{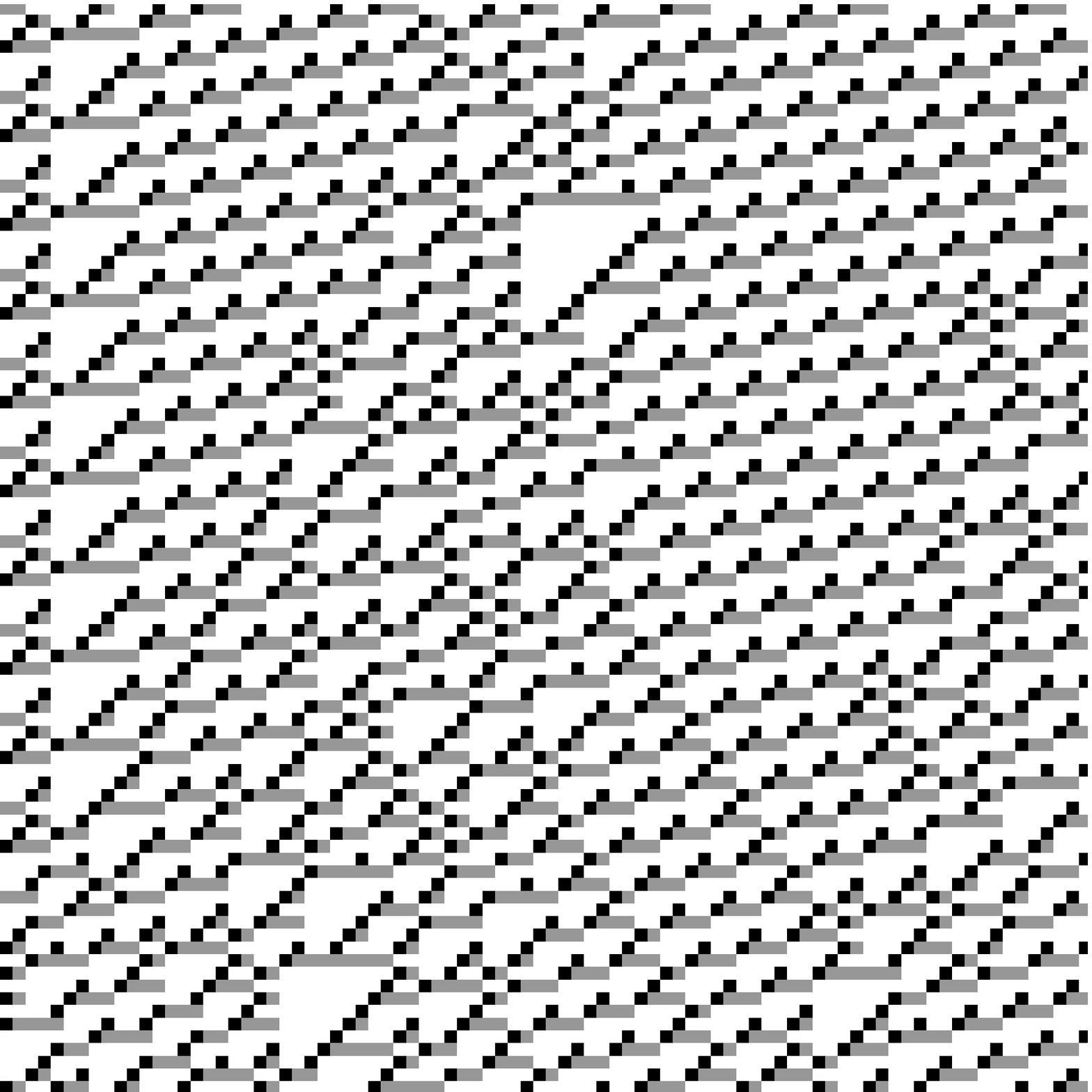}}}
	\subfigure[]{\fbox{\label{fig:110teApp1-1-pos}\includegraphics[width=0.23\textwidth]{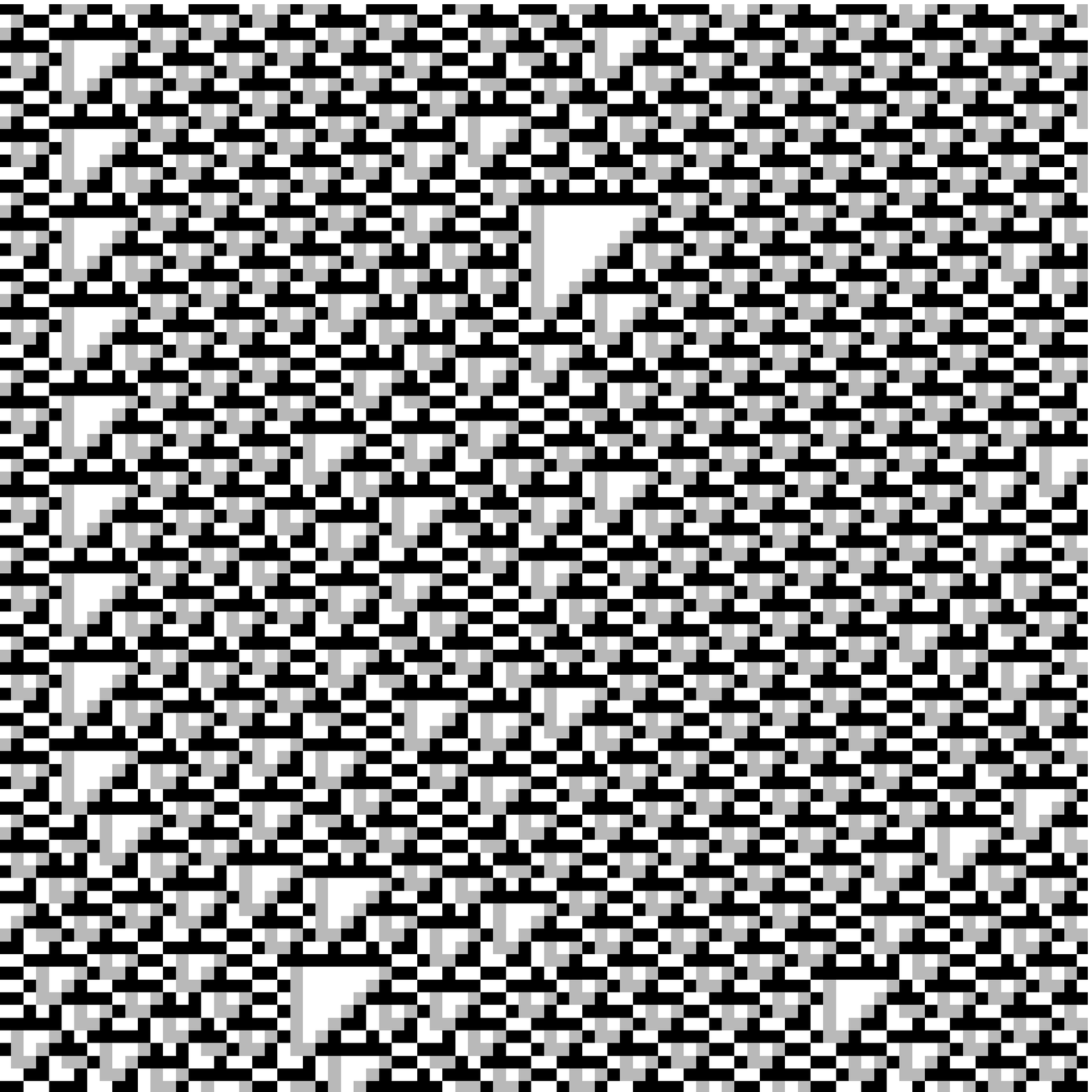}}}
	\caption{\label{fig:mutualK1}\optColourCaptionPrefix Base comparison metrics incapable of quantifying local information transfer (one cell to the right). Application to raw states of ECA Rule 110 shown in \subref{fig:110raw-base} (86 time steps displayed for 86 cells, time increases down the page for all CA plots):
		\subref{fig:110mi1-pos} Local (time-lagged) mutual information $m(i,j=1,n)$, positive values only, (all figures scaled with 16 colors) with max. 0.48 bits (\maxMagColour), min. 0.00 bits (white);
		\subref{fig:110teComp1-1} Local complete transfer entropy $t_c(i,j = 1,n,k=1)$, max. 1.28 bits (\maxMagColour), min. 0.00 bits (white);
		\subref{fig:110teApp1-1-pos} Local apparent transfer entropy $t(i,j = 1,n,k=1)$, positive values only, max. 0.67 bits (\maxMagColour), min. 0.00 bits (white).
	}
\end{figure}

For rule 110 the raw states of a sample CA run are displayed in Fig.~\ref{fig:110raw-base} (all figures were generated using modifications to \cite{woj02}). As base cases we measured (time-lagged) local mutual information $m(i,j,n)$, and local apparent and complete transfer entropies with the default value of $k=1$: $t(i,j,n,k = 1)$ and $t_c(i,j,n,k = 1)$. The base comparison case of local mutual information is analogous to that with globally averaged measures in \cite{schr00}, yet the local profiles yield a more detailed contrast here than averages do. Note that $k=1$ is the only value used in \cite{schr00} (in less coupled systems) and the later applications of the transfer entropy in \cite{ay06,lung06,bert06a}). The local profiles generated with $j=1$ (i.e. one cell to the right per unit time) for these base cases are shown in Fig.~\ref{fig:mutualK1}.
These measures are unable however to distinguish gliders from the background here with any more clarity than the raw CA plot itself. (The negative components of $m(i,j,n)$ and $t(i,j,n,k = 1)$, not shown, are similarly unhelpful). These basic metrics were also unsuccessful with other values of $j$ and with other CA rules; this provides explicit demonstration that they are not useful as measures of information transfer in complex systems.

\subsection{\label{glidersHighlighted}Gliders as dominant information transfer agents}
\begin{figure}
	\subfigure[]{\fbox{\label{fig:110raw-k6}\includegraphics[width=0.23\textwidth]{110-raw}}}
	\subfigure[]{\fbox{\label{fig:110teSum-6}\includegraphics[width=0.23\textwidth]{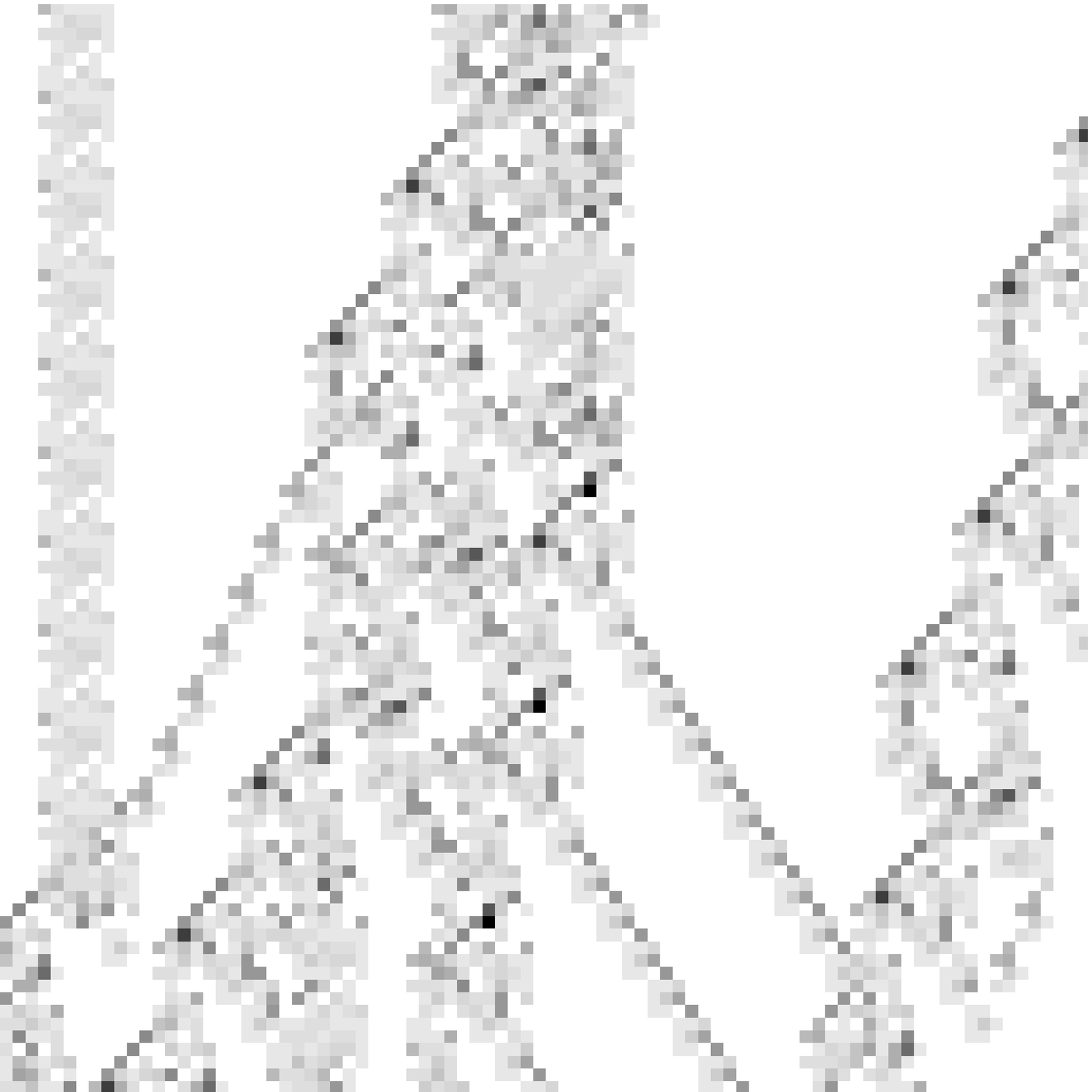}}}
	\subfigure[]{\fbox{\label{fig:110te1-6}\includegraphics[width=0.23\textwidth]{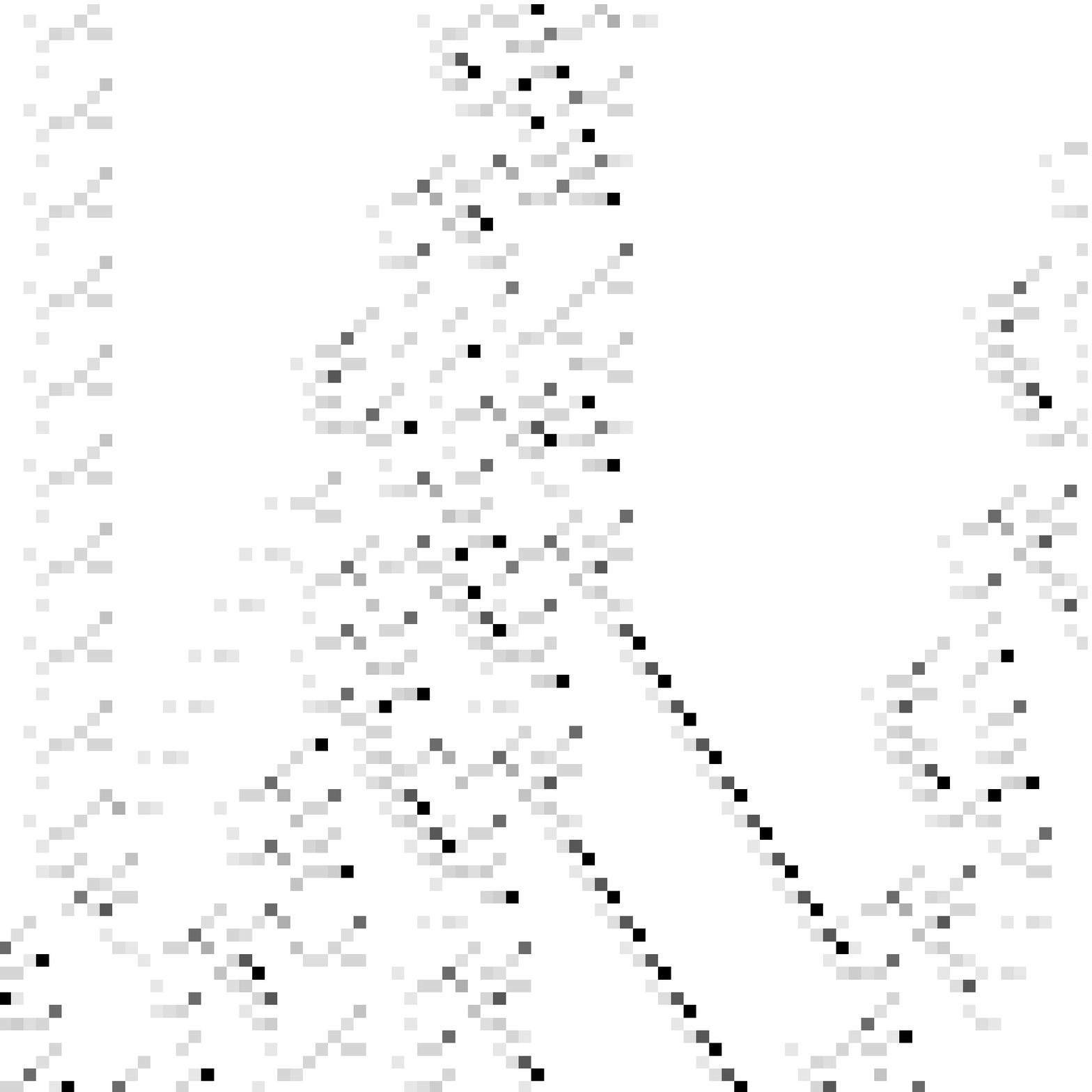}}}
	\subfigure[]{\fbox{\label{fig:110te-1-6}\includegraphics[width=0.23\textwidth]{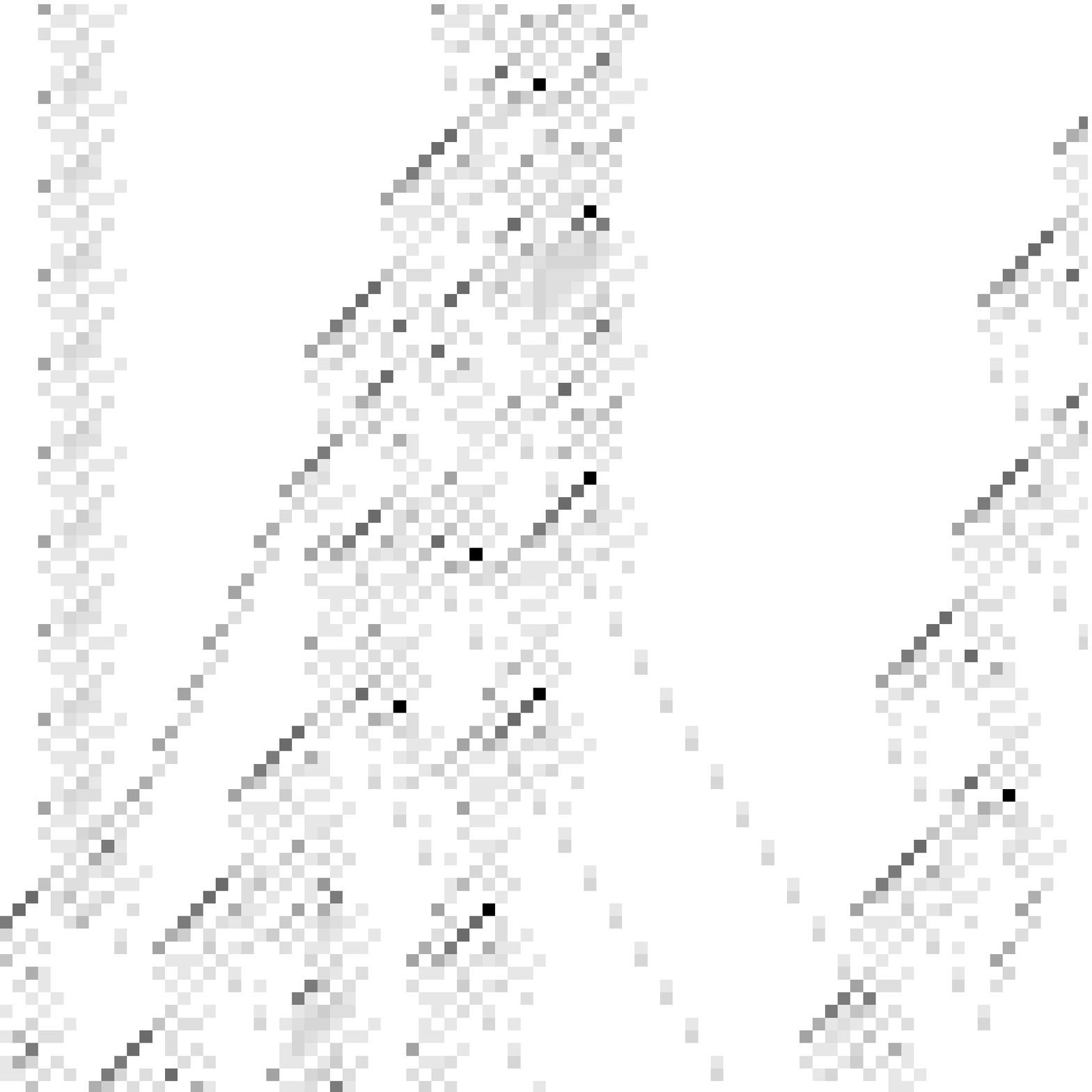}}}
	\caption{\label{fig:infoTx110K6}\optColourCaptionPrefix Local transfer entropy with $k = 6$ highlights gliders. Application to raw states of ECA Rule 110 in \subref{fig:110raw-k6} (86 time steps for 86 cells):
		\subref{fig:110teSum-6} Summed local complete transfer entropy profile $t_{sc}(i,n,k = 6)$, max. 8.22 bits (\maxMagColour), min. 0.00 bits (white);
		\subref{fig:110te1-6} Local complete transfer entropy $t_{c}(i,j=1,n,k=6)$ (one cell to the right), max. 4.95 bits, min. 0.00 bits;
		\subref{fig:110te-1-6} Local complete transfer entropy $t_{c}(i,j=-1,n,k=6)$ (one cell to the left), max. 6.72 bits, min. 0.00 bits.}
\end{figure}

Experimentally, we find our expectation of gliders being highlighted as dominant information transfer against the domain once $k \geq 6$ for ECA rule 110 (for both the complete and apparent metric, in both channels $j = 1$ and $-1$). Fig.~\ref{fig:infoTx110K6} displays the local complete transfer entropy profiles computed here using $k=6$ (we return to examine the apparent metric in Section \ref{compApparent}).
Note that higher values of local complete transfer entropy are attributed by each measure to the gliders moving in the \textit{same} macroscopic direction of motion as the direction of information transfer being measured, as is expected from such measures. 
Also, the summed local complete transfer in Fig.~\ref{fig:110teSum-6} gives a filtered plot very similar to that found for rule 110 using other techniques (see \cite{wue99, sha06}). Simply relying on the average transfer entropy values does not provide us these details (see Section \ref{averaged}).

\begin{figure}
	\subfigure[]{\fbox{\label{fig:110raw-closeup}\includegraphics[width=0.15\textwidth]{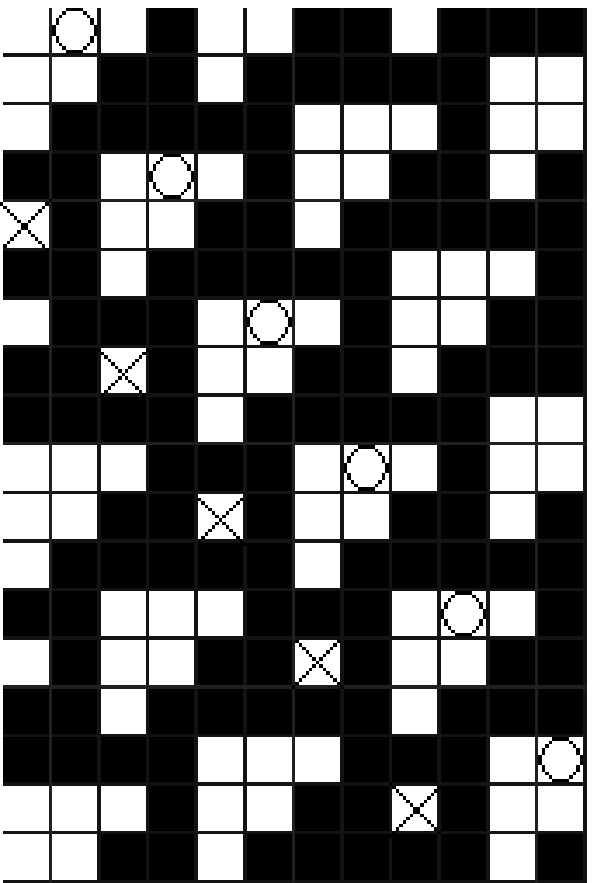}}}
	\subfigure[]{\fbox{\label{fig:110teSum6}\includegraphics[width=0.15\textwidth]{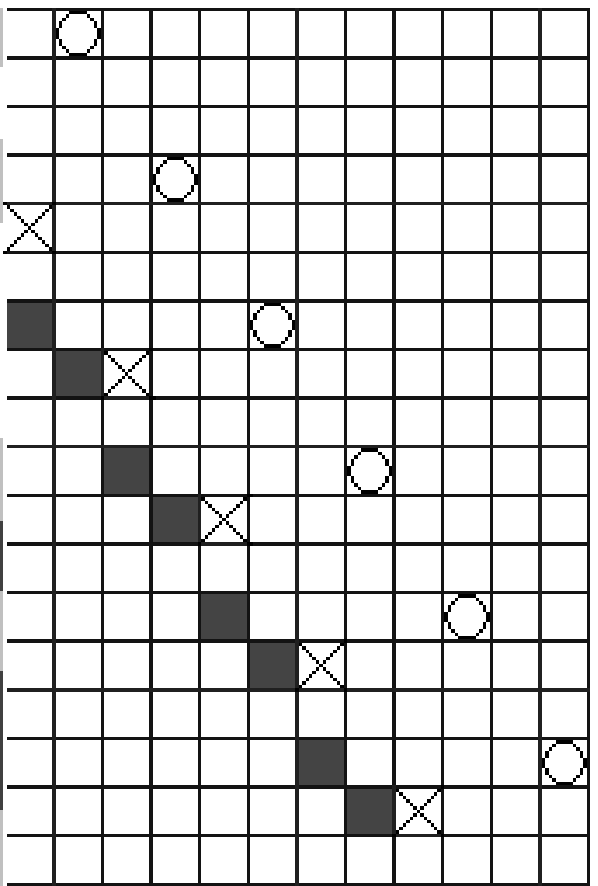}}}
	\subfigure[]{\fbox{\label{fig:110te-1-6-neg}\includegraphics[width=0.15\textwidth]{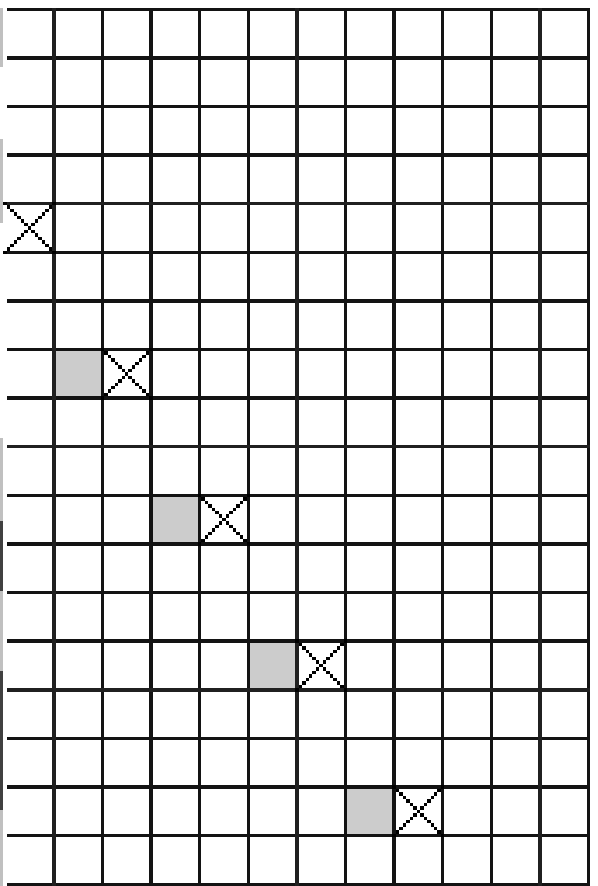}}}
	\caption{\label{fig:infoTx110Focus} Close-up example of a glider in ECA Rule 110 (x's and o's used only for visual alignment). 18 time steps displayed for 12 cells:
		\subref{fig:110raw-closeup} Raw CA;
		\subref{fig:110teSum6} Local complete transfer entropy $t_{c}(i,j=1,n,k=6)$ (one cell to the right), maxima in view 4.70 bits (gray), minima 0.00 bits (white);
		\subref{fig:110te-1-6-neg} Local \textit{apparent} transfer entropy $t(i,j=-1,n,k=6)$ (one cell to the left), negative values only, minima in view -2.04 bits (gray), maxima 0.00 bits (white).}
\end{figure}

Fig.~\ref{fig:110raw-closeup} displays a close-up example of a right moving glider in ECA rule 110, which application of the local complete transfer entropy in Fig.~\ref{fig:110teSum6} reveals is composed of a repeating series of two consecutive information transfers to the right followed by a pause. Although one may initially suggest that the glider structure includes the points marked ``x", careful consideration of exactly where a source can add information to that contained in the past of the domain suggests otherwise. Consider the point one cell to the left of those marked ``x", the second of the two consecutive transfers to the right. To compute $t_{c}(i,j=1,n+1,k=6)$ (one cell to the right) at this point, we first compute $p(x_{i,n+1}|x^{(k=6)}_{i,n},x_{i-1,n},d_{i,j=1,r,n}) = 1.0$ (since the system is deterministic) and $p(x_{i,n+1}|x^{(k=6)}_{i,n},d_{i,j=1,r,n}) = 0.038$. The local transfer entropy will be high here because the probability of observing the actual next state of the destination is much \textit{higher} when the source is taken into account than when it is not; correspondingly using Eq.~(\ref{eq:teLocalComplete}) we have $t_{c}(i,j=1,n+1,k=6) = 4.7$ bits at this point. The points marked ``x" are effectively predictable from the temporal pattern of the preceding domain however, and so do not contain significant information transfer.
Interestingly, the points containing significant information transfer are not necessarily the same as those selected as particles by other filtering methods; e.g. finite state transducers (using left to right scanning by convention \cite{han97}) would identify points two cells to the \textit{right} of those marked ``x" as part of the glider.

To understand why $k \geq 6$ was useful in this case, we consider an \textit{infinite} temporally periodic domain, with period say \textit{p}. (This serves as an extension of the demonstration in \cite{schr00} of zero average transfer in a lattice of spatial and temporal period 2 using $k = 1$ to a domain of arbitrary period). For the time-series of a single cell there, the number of states an observer must examine to have enough information to determine the next state is limited by the period $p$ (as per the synchronization time $\tau$ in \cite{feld04}). Local transfer entropy measurements with $k \geq p - 1$ would therefore not detect any additional information from the neighbors about the next state of the destination than is already contained in these \textit{k} previous states (correctly inferring zero transfer). Using $k < p - 1$ on the other hand may attribute the non-traveling self-influence of the destination to the source. Taking $k \geq p-1$ provides a \textit{sufficient} (Markovian) condition for eliminating this non-traveling information in an infinite periodic domain, rather than requiring the full asymptote $k \rightarrow \infty$. Establishing a \textit{minimal} condition is related to the synchronization time $\tau$ for the entropy rate \cite{feld04}, though is slightly more complicated here because we need to consider the source cell.

However, a minimal correct value for $k$ does not exist for a given system with bidirectional communication \textit{in general}. The above argument was only applicable for domains which are periodic and \textit{infinite}, and the existence of any gliders prevents a periodic domain from being infinite. Where the history of a given destination includes encountering gliders at some point, this partial knowledge of nearby glider activity is an important component in the probability distribution of the next state of that destination. Yet there is no limit on how far into the future a previous glider encounter may influence the states of a destination (because of the system's capacity for bidirectional communication). That is to say, there is no Markovian condition for eliminating the non-traveling information \textit{in general} in such systems; as such the limit $k \rightarrow \infty$ should be taken in measuring the transfer entropy. While using only the condition $k \geq p - 1$ is not completely correct, it will eliminate the non-traveling information in the domain pertaining to the periodic structure only. Where this part is dominant in the domain, as in for ECA rule 110 here, the gliders are likely to be highlighted against the periodic domain with $k \geq p - 1$. (This could be considered a rule of thumb for determining a minimum useful $k$).

\begin{figure}
	\subfigure[]{\fbox{\label{fig:110teSum-16}\includegraphics[width=0.23\textwidth]{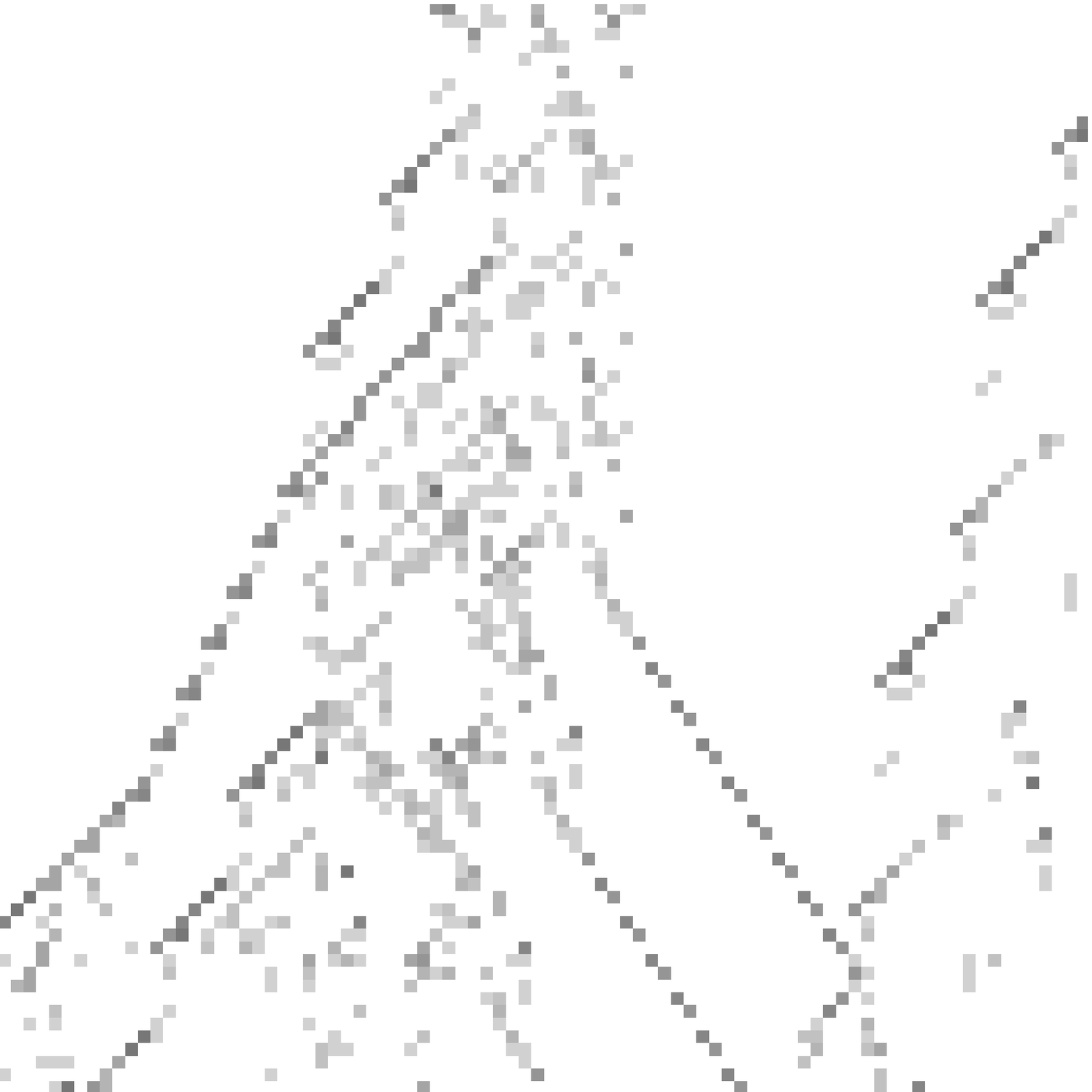}}}
	\subfigure[]{\fbox{\label{fig:110te1-16}\includegraphics[width=0.23\textwidth]{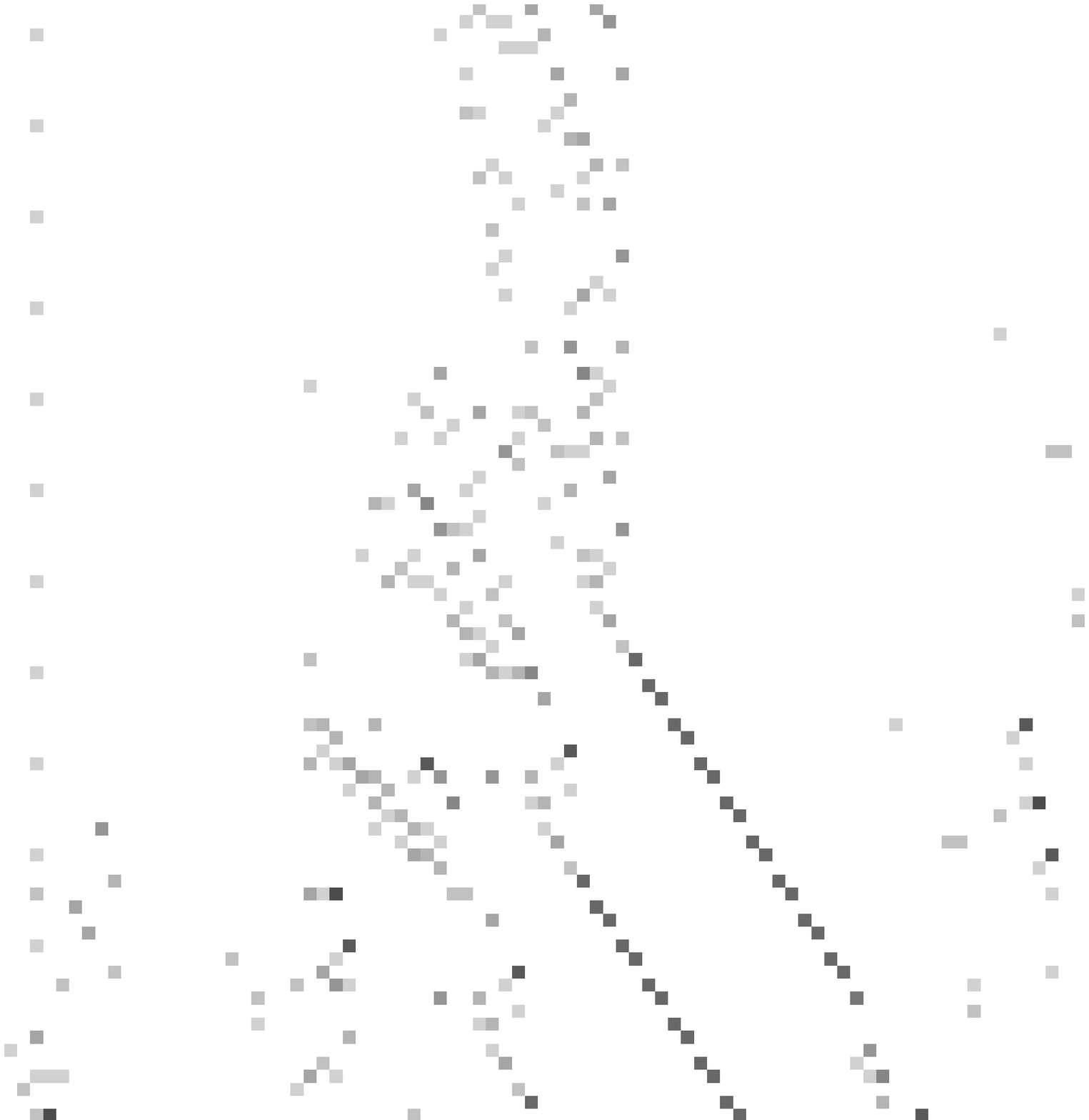}}}
	\subfigure[]{\fbox{\label{fig:110te-1-16}\includegraphics[width=0.23\textwidth]{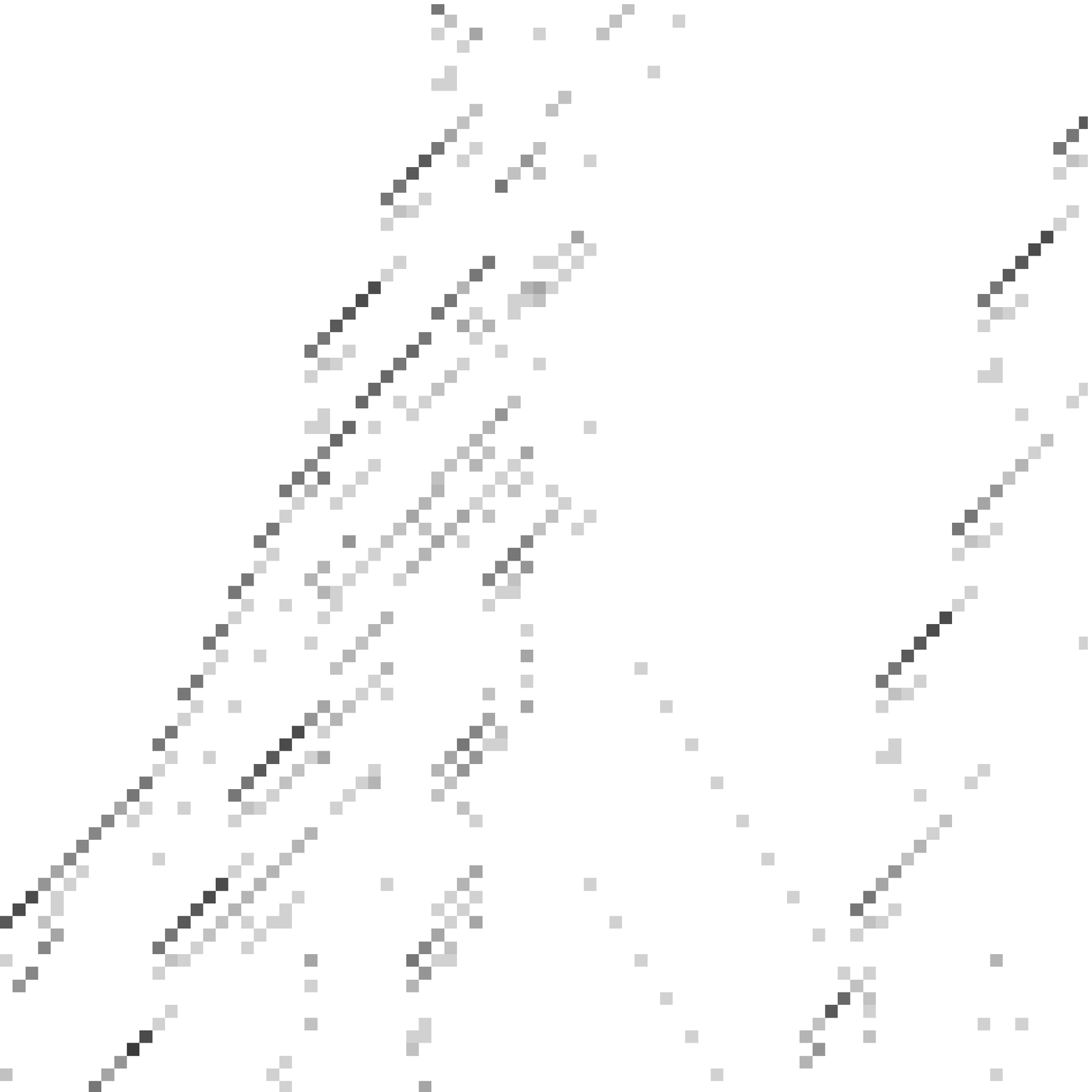}}}
	\subfigure[]{\fbox{\label{fig:110teApp-1-16-pos}\includegraphics[width=0.23\textwidth]{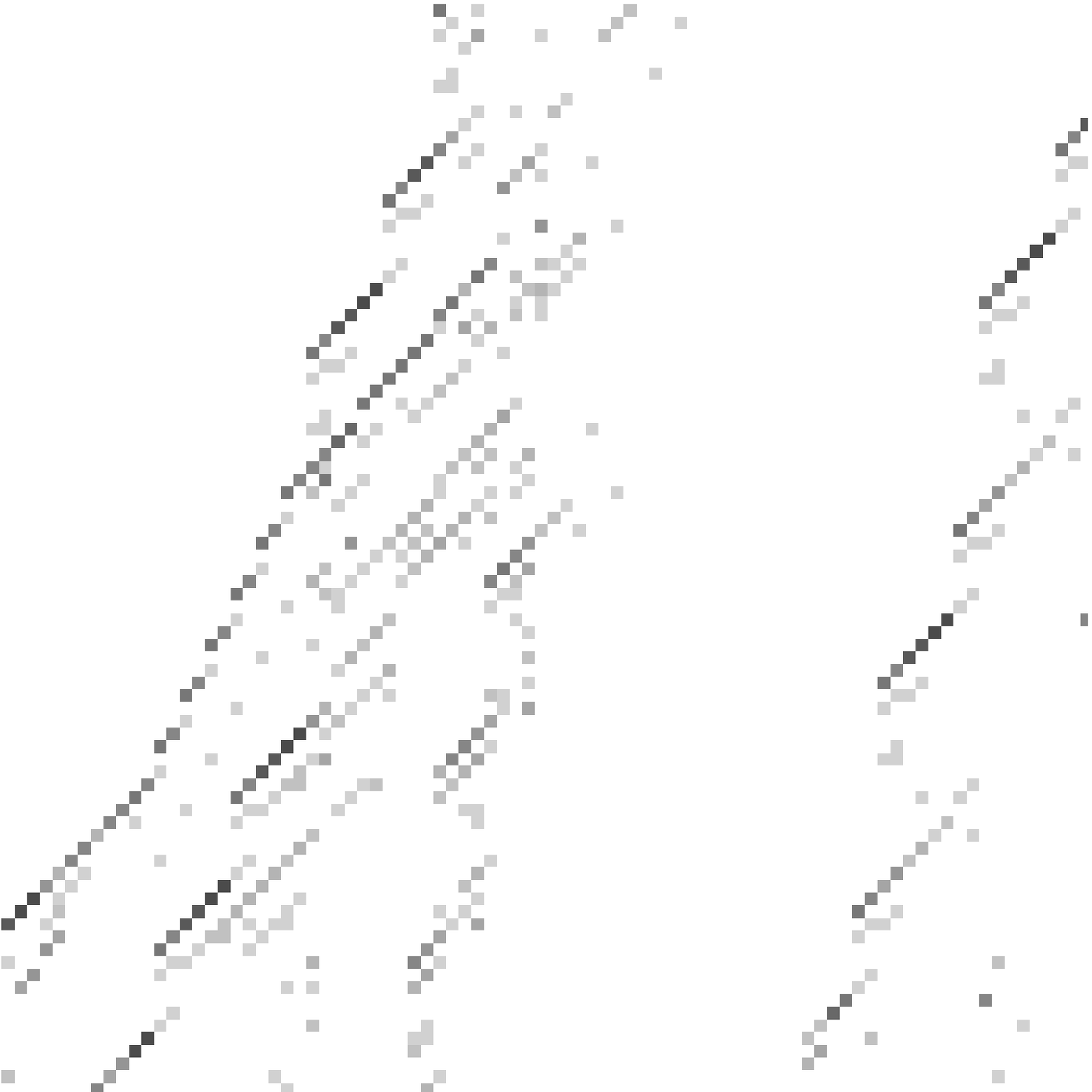}}}
	\caption{\label{fig:infoTx110K16}\optColourCaptionPrefix Estimating local transfer entropy profiles for $k \rightarrow \infty$ with $k=16$ for the raw states of ECA Rule 110 in Fig.~\ref{fig:110raw-k6} (86 time steps for 86 cells):
		\subref{fig:110teSum-16} Summed local complete transfer entropy profile $t_{sc}(i,n,k = 16)$, max. 14.7 bits (\maxMagColour), min. 0.00 bits (white);
		\subref{fig:110te1-16} Local complete transfer entropy $t_{c}(i,j=1,n,k=16)$ (one cell to the right), max. 9.99 bits, min. 0.00 bits;
		\subref{fig:110te-1-16} Local complete transfer entropy $t_{c}(i,j=-1,n,k=16)$ (one cell to the left), max. 10.1 bits, min. 0.00 bits;
		\subref{fig:110teApp-1-16-pos} Local \textit{apparent} transfer entropy $t(i,j=-1,n,k=16)$ (one cell to the left), positive values only, max. 10.4 bits, min. 0.00 bits.
	}
\end{figure}

While the results for $k=6$ visually correlate with previous filtering work, using the limit $k \rightarrow \infty$ would be more correct. Achieving this limit is not computationally feasible, but reasonable estimates of the probability distributions can be made: Fig.~\ref{fig:infoTx110K16} displays the local complete transfer entropy profiles computed for ECA rule 110 using $k=16$.
These plots highlight information transfer almost exclusively now in the direction of the macroscopic glider motion, which is even more closely aligned with our expectations than was seen for $k=6$. Importantly, much less of the gliders are highlighted than for $k = 6$ or other techniques, and the larger values of transfer entropy are concentrated around the leading time-edges of the gliders. This suggests that the leading glider edges determine much of the following dynamics which then comprise mainly non-traveling information.
Note also that the ``vertical" glider (at the left of Fig.~\ref{fig:110teSum-6}, with spatial velocity zero) is not highlighted now. Its cell states are effectively predictable from their past, observable once $k$ becomes greater than its vertical period.

Another interesting effect of the existence of gliders is that the next state of a cell in the domain is not completely determined by its periodic history. The neighboring information sources have the capability to \textit{add} information about the next state of that destination, by signaling whether a glider is incoming or not. That is to say, it is possible to measure a non-zero information transfer inside finite domains, effectively indicating the \textit{absence} of a glider (i.e. that the domain shall continue). For ECA rule 110 in Fig.~\ref{fig:infoTx110K6}, we do in fact measure small but non-zero information transfer at certain points in the periodic background domain (small enough to appear to be zero). These values tend to be stronger in the wake of real gliders: since gliders are often followed by others, there is a stronger indication of their absence. Consider the points in the periodic domain marked by ``o" in Fig.~\ref{fig:infoTx110Focus}: these have the same history as the previously discussed points of high information transfer; their neighborhood (excluding the source on the left) is also the same. Here, we compute $p(x_{i,n+1}|x^{(k=6)}_{i,n},x_{i-1,n},d_{i,j=1,r,n}) = 1.0$
and $p(x_{i,n+1}|x^{(k=6)}_{i,n},d_{i,j=1,r,n}) = 0.96$: the probability of observing the actual next state of the destination becomes slightly higher when the information source on the left is taken into account. As such, we have $t_{c}(i,j=1,n+1,k=6) = 0.057$ bits to the right at this point, demonstrating the possibility for small non-zero information transfer in the periodic domain.
This effect occurs for both the complete and apparent measures and is not a finite $k$ effect.

Also, note in Fig.~\ref{fig:infoTx110K16} there is some information transfer in the orthogonal direction for each glider. Some is expected to vanish as $k \rightarrow \infty$, yet some will remain for a similar reason to the non-zero transfer in domains, i.e. considering the source does add information about the next state of the destination.
Importantly, this orthogonal transfer is not as significant as that in the macroscopic glider direction in terms of magnitude and coherence.

Given these effects, we describe gliders as the \textit{dominant}, as opposed to the only, information transfer agents here. (These findings have also been verified for ECA rule 54, another complex rule containing gliders.)
While these profiles appear similar to other filtering work in some respects, it is only local transfer entropy profiles that provide quantitative evidence that gliders are the dominant information transfer agents in CAs.

\subsection{\label{domainWalls}Domain walls as dominant information transfer agents}
\begin{figure}
	\subfigure[]{\fbox{\label{fig:18raw}\includegraphics[width=0.23\textwidth]{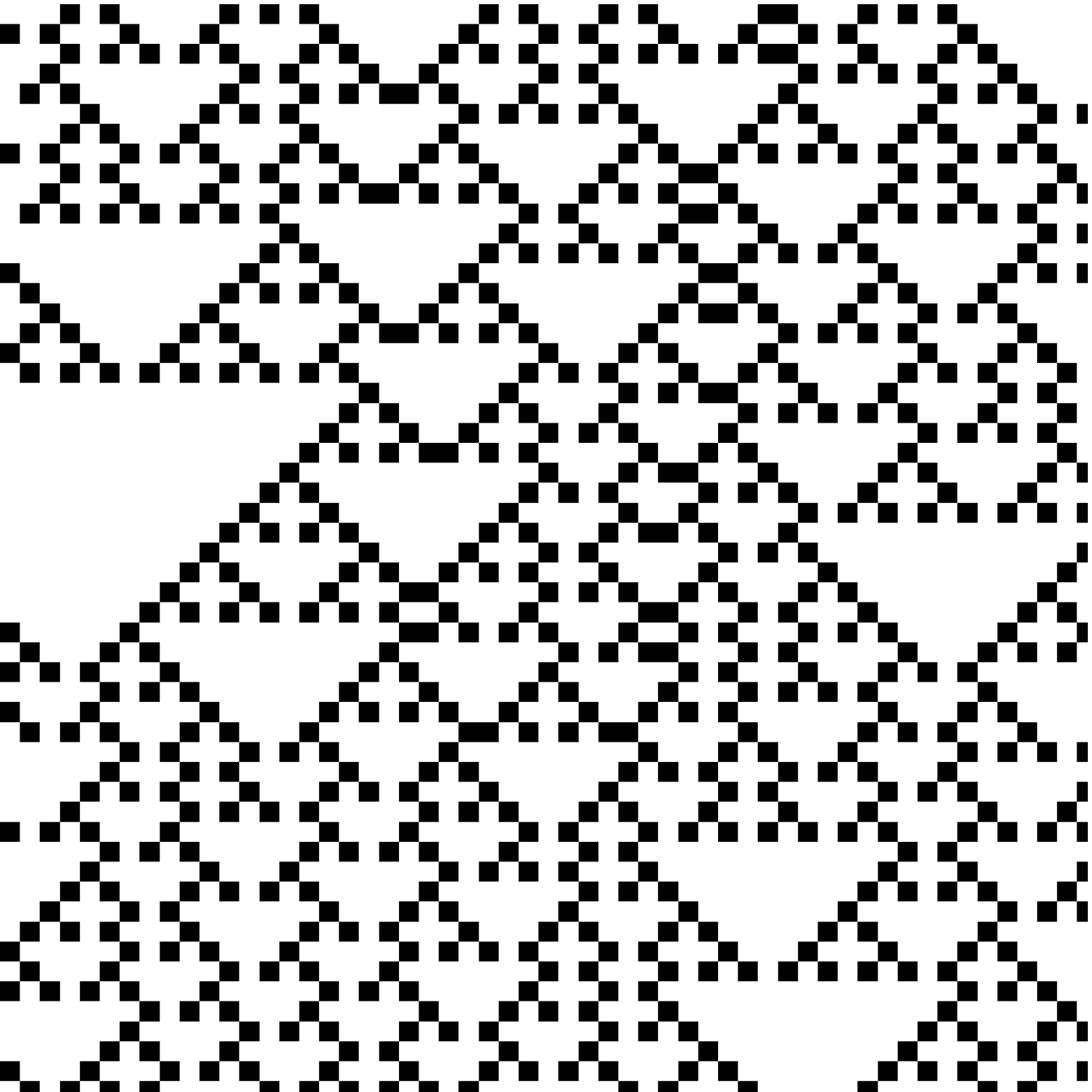}}}
	\subfigure[]{\fbox{\label{fig:18teSum16}\includegraphics[width=0.23\textwidth]{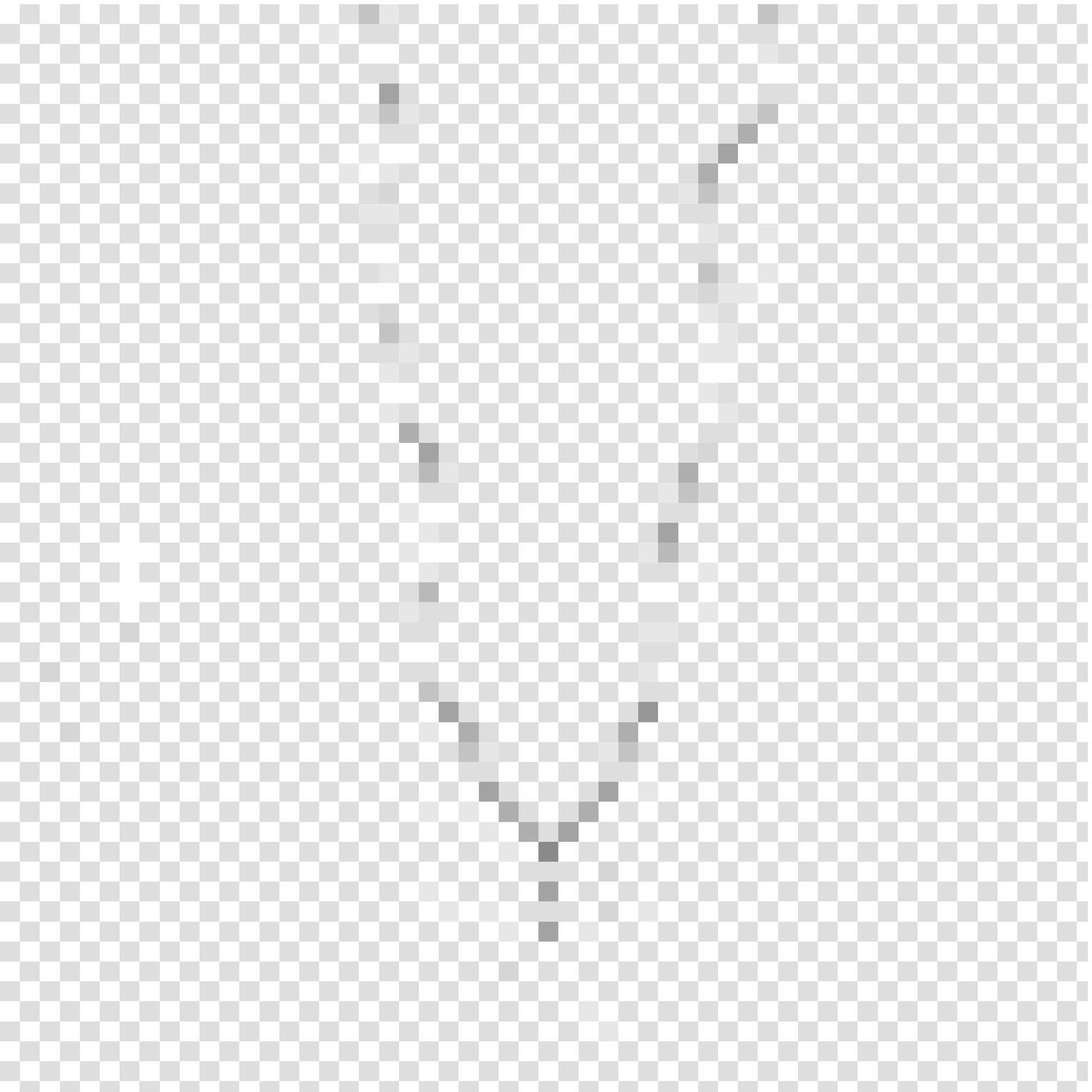}}}
	\subfigure[]{\fbox{\label{fig:18te1-16}\includegraphics[width=0.23\textwidth]{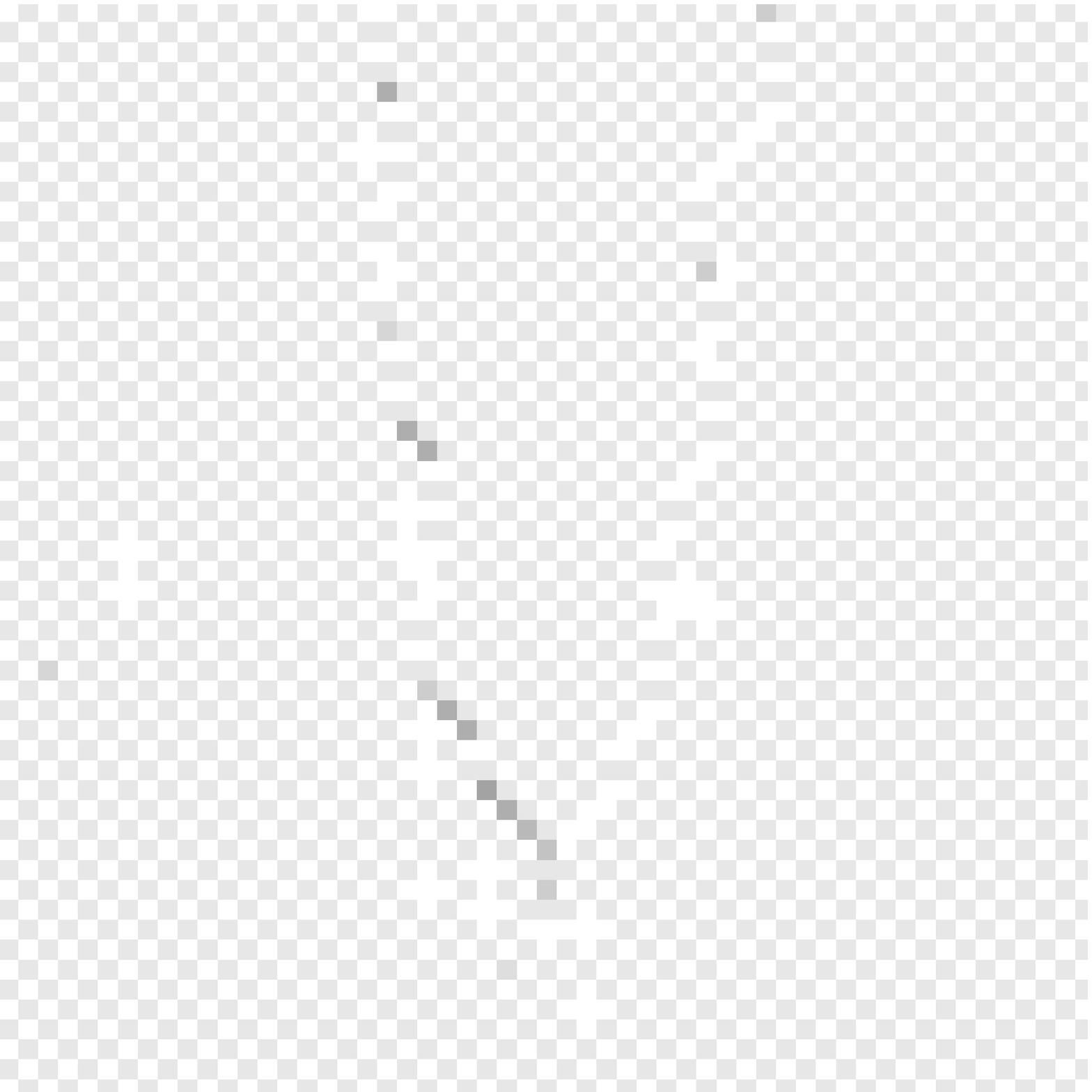}}}
	\subfigure[]{\fbox{\label{fig:18teApp1-16-Pos}\includegraphics[width=0.23\textwidth]{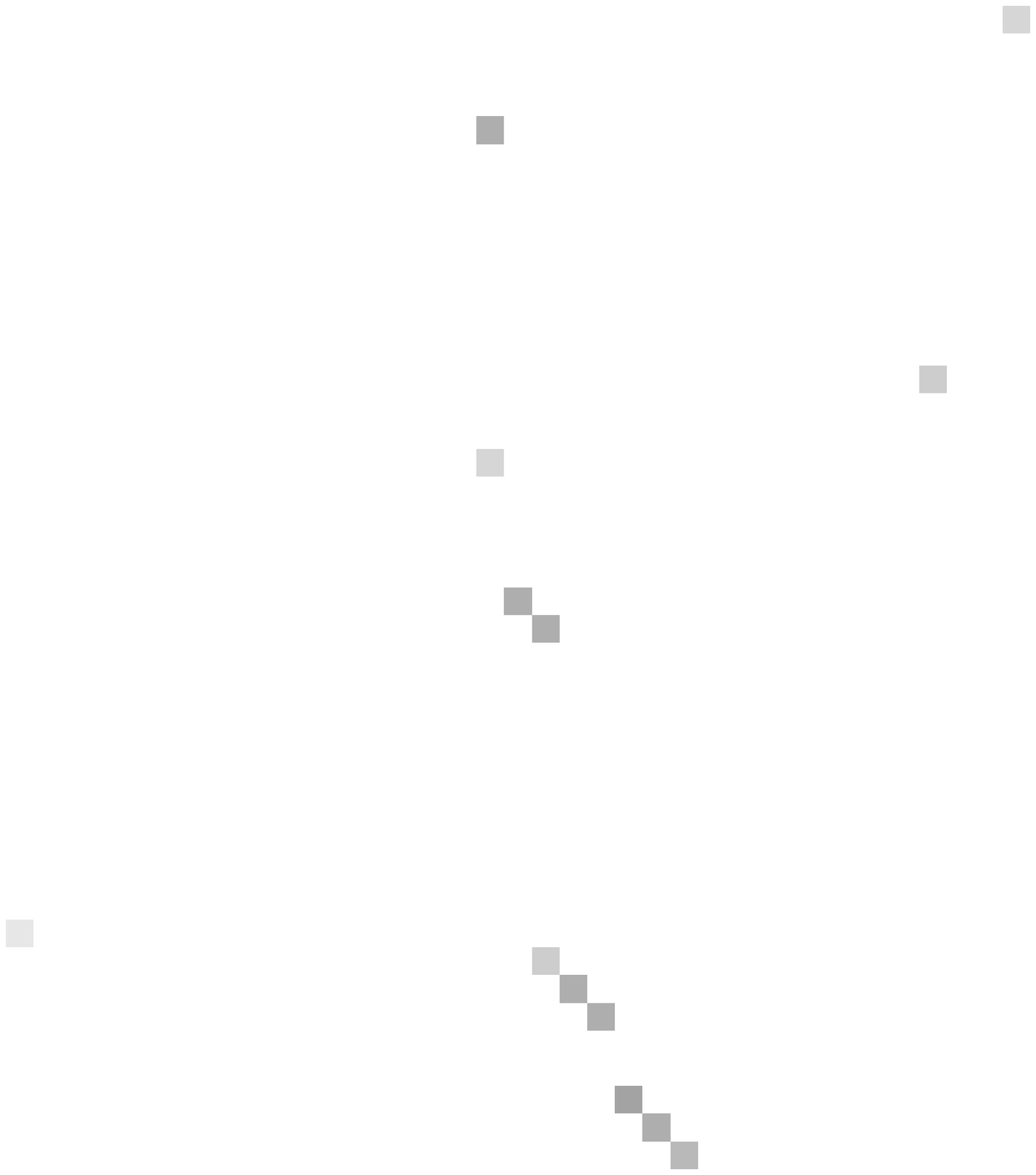}}}
	\caption{\label{fig:infoTx18}\optColourCaptionPrefix Local transfer entropy profiles for raw states of ECA Rule 18 in \subref{fig:18raw} (55 time steps for 55 cells displayed) highlight domain walls:
		\subref{fig:18teSum16} Summed local complete transfer entropy profile $t_{sc}(i,n,k = 16)$, max. 13.5 bits (\maxMagColour), min. 0.00 bits (white);
		\subref{fig:18te1-16} Local complete transfer entropy $t_{c}(i,j=1,n,k=16)$ (one cell to the right), max. 14.9 bits, min. 0.00 bits;
		\subref{fig:18teApp1-16-Pos} Local \textit{apparent} transfer entropy $t(i,j=1,n,k=16)$ (one cell to the right), positive values only, max. 11.9 bits, min. 0.00 bits;
	}
\end{figure}

We also investigated ECA rule 18, known to contain domain walls against the background. Application of local complete transfer entropy to the sample run in Fig.~\ref{fig:18raw} highlights the domain walls as containing strong information transfer in each channel (e.g. see $t_c(i,j=1,n,k=16)$ in Fig.~\ref{fig:18te1-16}). A full picture is given by the summed profile in Fig.~\ref{fig:18teSum16}: as expected, our results quantitatively confirm the domain walls as dominant information transfer agents against the domain. We have observed similar results for ECA rule 146.

Importantly, the domain contains a significant level of information transfer here. In fact, there is a pattern to the transfer in the domain of spatial and temporal period 2 which corresponds very well to the period-2 spatial $\epsilon$-machine generated to recognize the domain of rule 18 in \cite{han92}. Every second site of the domain in the raw CA is a ``0", and the alternate site is either a ``0" or a ``1" (depending on the neighborhood configuration).
At every second site with the ``0" values, there is vanishing local complete information transfer (for either incoming channel $j = 1$ or $-1$) because the state of the cell is completely predictable from this temporal periodic pattern in its past. At the alternate sites, the local complete information transfer is approximately 1 bit from both incoming channels $j = 1$ and $-1$ (by limited inspection the measurements were between 0.96 and 1.04 bits with $k=16$). At these points, (in an infinite domain) both alternative next states are equally likely (in the context of the destination's past and the rest of the CA neighborhood) before considering the source; when it is considered, the next state is determined and 1 bit of information is added.

Domain walls involve the meeting of two domains which are \textit{out of phase}: motion of the wall can be viewed as one domain intruding into the other. At such points, we observe high transfer entropy in the direction of movement because the information source (as part of the intruding domain) adds much information about the next state of the destination that was not in the destination's past or the rest of the CA neighborhood.
This highlighting of the domain walls is somewhat similar to that produced by other filtering techniques, although an important distinction to \cite{han92,wue99,hel04} is that this technique highlights the domain wall areas as only being a single cell wide: as described above, a single cell width is all that is required to explain the meeting of two domains of rule 18 from a temporal perspective.


We also applied these measures to ECA rule 22 (not shown), plots of whose raw states appears similar to rule 18 at first glance. However, this rule has not been found to contain structure such as domain walls \cite{sha06}. Similar to those results, local transfer entropy measures significant information transfer at many points in the CA, but does not find any coherent structure to this transfer.

\subsection{\label{compApparent}Apparent transfer entropy}
Profiles generated with the local apparent transfer entropy contain many of the same features as those for the complete metric: gliders and domain walls are highlighted as the dominant information transfer agents in their direction of motion; large values of $k$ are required to reasonably approximate the probability distribution functions; and non-zero information transfer is still possible in domains and in orthogonal directions to macroscopic glider motion.

For ECA rule 110, Fig.~\ref{fig:110teApp-1-16-pos} displays the positive values for $t(i,j=-1,n,k=16)$ (one cell to the left), which appears almost identical to the corresponding profile for the complete metric in Fig.~\ref{fig:110te-1-16}. The summed apparent profile (not shown) is also very similar to the summed complete profile in Fig.~\ref{fig:110teSum-16}. A major distinction is observed however when examining negative values for the apparent profiles: when measured in a directional orthogonal to macroscopic glider motion, it can report \textit{negative} as well as positive values (see Fig.~\ref{fig:110te-1-6-neg}). Negative values occurs where the source, still part of the domain, is misleading about the next state of the destination.

As an example, consider the glider in Fig.~\ref{fig:110raw-closeup}. At the positions to the left of those marked ``x", we confirm a strong positive value for the local apparent transfer entropy $t(i,j=1,n+1,k=6)$ (2.65 bits), as per the complete metric. However, Fig.~\ref{fig:110te-1-6-neg} displays large negative values of $t(i,j=-1,n+1,k=6)$ (the orthogonal channel to glider motion) at these same positions. There we compute $p ( x_{i,n+1}|x^{(k=6)}_{i,n},x_{i-1,n} ) = 0.038$ and $p ( x_{i,n+1}|x^{(k=6)}_{i,n} ) = 0.16$. The local apparent transfer entropy negative here because the probability of observing the actual next state of the destination is much \textit{lower} when the source on the right is taken into account than when it is not (i.e. the source is \textit{misleading}). As such, Eq.~(\ref{eq:teLocal}) gives $t(i,j=-1,n+1,k=6) = -2.05$ bits at this point. Compare this to the complete metric for this channel, $t_c(i,j=-1,n+1,k=6)$, which measures 0.00 bits here because the source at the right (still in the domain) cannot add any information not contained in the other neighbor (which drives the glider). Note that the local apparent transfer entropy in the direction of glider motion was more informative than that in the orthogonal direction was misleading.
Also, note that negative values of the local metric are \textit{not} found for the orthogonal direction at every point in the glider.

Another distinction is observed for ECA rule 18. As expected, the apparent metric identifies high positive transfer entropy in the direction of domain wall motion (see Fig.~\ref{fig:18teApp1-16-Pos} for the $j=1$ channel), and negative transfer entropy in the orthogonal direction to domain wall motion (not shown). However, the apparent metric finds vanishing transfer entropy throughout the domain (for both channels $j=-1$ and $1$), in stark contrast to the periodic pattern found with the complete metric. At every second site with the ``0" values, the state of the destination is completely predictable from its past, so we have $t = 0$ bits as for $t_c$. However, at the alternate sites both possible next states are equally likely in the context of the destination's history and remain so when considering the source: as such we find $t=0$. It is only when including the rest of the neighborhood in the context (with the complete metric) that one observes the source to be adding 1 bit of information. This example brings to mind discussion on the nature of information transfer in complex versus chaotic dynamics \cite{sole01,mira95,lang90,coff98} and suggests that perhaps in chaotic dynamics, where many sources influence outcomes in a non-coherent manner, the complete metric may indicate large information transfer whereas the apparent metric does not (because other sources obscure the contribution of the source under consideration).

The apparent and complete metrics are clearly capable of producing different insights under certain circumstances, and both viewpoints are valuable. We are currently investigating an application of the apparent transfer entropy in combination with a measure of information storage to identify information modification \cite{liz07b}.

\subsection{\label{averaged}Averaged transfer entropies}
\begin{figure}
	\includegraphics[width=0.48\textwidth]{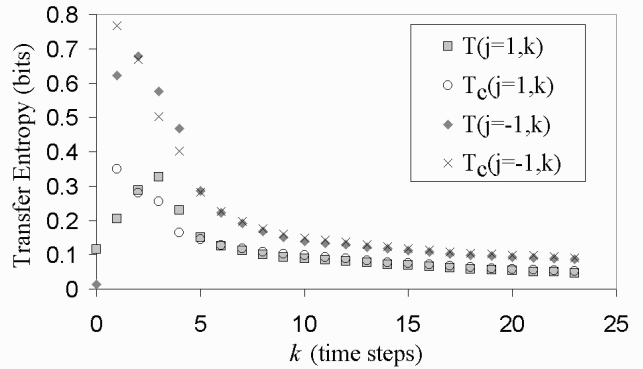}
	\caption{\label{fig:functionOfK} Average transfer entropies versus conditioning length $k$, plotted for complete and apparent transfer entropies in channels $j = 1$ and $-1$ in ECA rule 110.}
\end{figure}

We compute the averaged transfer metrics as a function of $k$ for ECA rule 110 in Fig.~\ref{fig:functionOfK} so as to check whether similar insights can be gained from this trend.
In fact, only limited insights are gained here. The average complete transfer entropies decrease with $k$: an increase is impossible because we condition out more of the information that appears to come from the source. The average apparent transfer entropy can show increases with $k$ however; this is possible with a three-term entropy \cite{mac03} where other information sources are not taken into account. None of these reach a limiting value for the extent of $k$ measured, suggesting again that $k \rightarrow \infty$ should be used. Realistically, $k$ is limited (e.g. to $k=16$ in previous sections) by the sample size so as to retain a sufficient number of observations per configuration to reasonably estimate the probability distribution functions.

The local metrics clearly reveal much about the information dynamics of a system that their averages do not.
In particular, these averages tell us nothing of the presence of glider particles, not to mention that they would be clearly highlighted once $k \geq 6$. Also, while the average apparent and complete metrics appear to be converging to a similar value in each channel, this belies their important distinctions discussed earlier.

\section{\label{conclusion}Conclusion}
We have presented a local formulation of the transfer entropy in order to characterize the information transfer into each spatiotemporal point in a complex system. Local transfer entropy presents insights that cannot be obtained using the averaged measure alone, in particular in providing these spatiotemporal information transfer profiles as an analytic tool. Importantly, the local transfer entropy allowed us to study the transfer entropy metric itself, including the importance of appropriate destination conditioning lengths $k$ (e.g. that using $k \rightarrow \infty$ is most correct), and to contrast the apparent and complete forms which were introduced here.

On applying the local transfer entropy to cellular automata, we demonstrated its utility as a valid filter for coherent structure. It is novel in comparison to other filtering methods previously presented for CAs. It provides continuous rather than discrete values (like \cite{hel04} and \cite{sha06}). It does not follow an arbitrary spatial preference (unlike \cite{hel04} and \cite{han92}) but rather the flow of time only. As described for local statistical complexity in \cite{sha06}, local transfer entropy does not require a new filter for every CA, but the probability distribution functions must be recalculated for every CA. Perhaps most importantly, it provides multiple views of information transfer in each generic channel or direction (which no other filters do), and also provides a combined view which matches many important features highlighted by other filters. Finally, it highlights subtly different parts of emergent structure to other filters, i.e.: the leading glider edges facilitating the information transfer; only the minimal part of domain walls necessary to identify them; the particles are identified as consisting of different points due to our temporal approach; and it does not highlight vertical gliders since they are not traveling information.

Most significantly, local transfer entropy provided the first quantitative support for the long-held conjecture that particles (both gliders and domain walls) are the information transfer agents in CAs. This is particularly important because of analogies between particles in CAs and coherent structure or hypothesized information transfer agents in physical systems, such as traveling localizations caused by dipole-dipole interactions in microtubules \cite{brown99} and in soliton dynamics \cite{park86}. This formulation of local transfer entropy is ready to be applied beyond CAs to systems such as these (and including stochastic systems), where it may prove similar conjectures about information transfer therein.

This result is important in bringing together the quantitative definition of information transfer (transfer entropy) with the popular understanding of the concept through widely-accepted instances (such as particles in CAs). The result therefore completes the establishment of transfer entropy as the appropriate measure for (predictive) information transfer in complex systems.
A comparison should be made with a localization of the ``information flow" metric \cite{ay06} in future work, in order to explore the differences between its causal perspective and the predictive or computational perspective of transfer entropy. In doing so, the limitations of the transfer entropy metric must be considered. These include that the transfer entropy should consider only causal information contributors as the source and as other information contributors to be conditioned on (in the complete metric). Considering non-causal sources (e.g. outside the neighborhood in CAs) has the potential to mistake correlation for information transfer, and conditioning on non-causal elements could cause information that was actually part of the transfer to be disregarded.

Finally, we are building on this investigation to describe local measures of information storage and modification also in a complete local framework for information dynamics in complex systems (see \cite{liz07b}).

\bibliography{LocalInfoTransfer}

\end{document}